\documentclass[10pt,a4paper]{article}

\usepackage{arxiv}
\usepackage{tikz}
\usepackage[utf8]{inputenc} 
\usepackage[T1]{fontenc}    
\usepackage{hyperref}       
\usepackage{url}            
\usepackage{booktabs}       
\usepackage{amsfonts}       
\usepackage{nicefrac}       
\usepackage{microtype}      
\usepackage{lipsum}

\usepackage{amssymb}
\usepackage{lineno,hyperref}
\usepackage{amsmath}
\usepackage{bm}
\usepackage{subcaption}
\usepackage{graphicx}
\usepackage{svg}
\usepackage{rotating}
\usepackage{lscape}
\usepackage{array,multirow}
\usepackage[title]{appendix}
\usepackage{float}
\usepackage{stmaryrd}
\usepackage{textcomp}
\usepackage{algorithm}
\usepackage{algpseudocode}
\usepackage{amsthm}
\usepackage{listings}
\usepackage{xcolor}
\usepackage{siunitx}

\usepackage{pifont}
\usepackage{lineno}

\usepackage{import}
\usepackage{physics}

\usetikzlibrary{positioning}

\usepackage[sort&compress]{natbib}
\title{The mixed Deep Energy Method for resolving concentration features in finite strain hyperelasticity}

\author{
  Jan Niklas Fuhg \\
  Sibley School of Mechanical and Aerospace Engineering \\
  Cornell University, 
   New York, USA \\
  \texttt{jf853@cornell.edu} \\

   \And
 Nikolaos Bouklas \\
  Sibley School of Mechanical and Aerospace Engineering\\
  Cornell University,
   New York, USA \\
  \texttt{nb589@cornell.edu} \\
  }
\DeclareMathOperator*{\argmin}{arg\,min}

\begin{document}
\maketitle

\begin{abstract}
The introduction of
Physics-informed Neural Networks (PINNs) has led to an increased interest in deep neural networks as universal approximators of PDEs in the solid mechanics community.
Recently, the Deep Energy Method (DEM) has been proposed. DEM is based on energy minimization principles, contrary to PINN which is based on the residual of the PDEs. A significant advantage of DEM, is that it requires the approximation of lower order derivatives compared to formulations that are based on strong form residuals.
However both DEM and classical PINN formulations struggle to resolve fine features of the stress and displacement fields, for example concentration features in solid mechanics applications.
We propose an extension to the Deep Energy Method (DEM) to resolve these features for finite strain hyperelasticity. The developed framework termed mixed Deep Energy Method (mDEM) introduces stress measures as an additional output of the NN to the recently introduced pure displacement formulation. Using this approach, Neumann boundary conditions are approximated more accurately and the accuracy around spatial features which are typically responsible for high concentrations is increased.
In order to make the proposed approach more versatile, we introduce a numerical integration scheme  based on Delaunay integration, which enables the mDEM framework  to be used for random training point position sets commonly needed for computational domains with stress concentrations, i.e. domains with holes, notches, etc.
We highlight the advantages of the proposed approach while showing the shortcomings of classical PINN and DEM formulations. The method is offering comparable results to Finite-Element Method (FEM) on the forward calculation of challenging computational experiments involving domains with fine geometric features and concentrated loads, but additionally offers unique capabilities for the solution of inverse problems and parameter estimation in the context of hyperelasticity.
\end{abstract}


\section{Introduction}
Solid mechanics studies the motion and deformation of solid materials when subjected to external forces. 
In the last years we have seen a rapid growth in the utilization of machine learning and data-driven techniques in computational solid mechanics. For example,
\cite{kirchdoerfer2016data} defined a computational data-driven approach that enforces constraints and conservation laws directly on experimental data. This data-driven approach has been extended to a wide range of directions such as inelastic problems \citep{eggersmann2019model} and also extended to a non-local setting \citep{gonzalez2019thermodynamically}. 
Machine learning has also been used to establish data-driven elastic and inelastic constitutive laws \citep{ibanez2017data, huang2020machine, fuhg2021modeldatadriven}. 
These ideas make use of existing infrastructures such as existing Finite-Element Method (FEM) solvers to solve the accompanying partial differential equations (PDEs). Additionally, machine learning techniques have been employed for the development of intrusive and non-intrusive Reduced Order Modeling (ROM) schemes for the accelerated solutions of PDEs \citep{kadeethum2021non,hernandez2021deep} following the offline-online paradigm, in which synthetic training data is often constructed from FEM solvers.

As an alternative approach, Neural Networks (NN) have recently gained a lot of attention in the solid mechanics community, and more broadly in the physic-based modeling community, for finding solutions of boundary value problems from scratch, due to their abilities as universal approximators \citep{hornik1989multilayer}. While the idea is not new and has already been proposed over 20 years ago \citep{lagaris1998artificial}, it has seen remarkable success recently due to advances in GPU hardware and neural network technologies ( e.g. deep learning libraries, stochastic gradient descent algorithms, etc). In this context Physics-informed Neural Networks (PINNS), first introduced in \cite{raissi2019physics}, have emerged as a deep learning algorithm for solving PDEs, based on automatic differentiation. The method has several advantages compared to classical numerical techniques with the major ones being that the algorithm is meshfree and is significantly easier to implement than traditional methods such as FEM while at the same time allows for the incorporation of constraints due to experimental observations and solution of inverse problems for parameter estimation. As PINN is an optimization-based solution technique, it also has the potential to be utilized in highly nonlinear problems where traditional solution techniques are not sufficient or necessarily robust. PINN has been extended into different directions such as variational problems (VPINN) \citep{kharazmi2021hp} or a subdomain-version (XPINN) which defines local neural networks in order to allow parallelization of the computation and also to resolve the solution in heterogeneous domains \citep{jagtap2020extended}.

From these ideas different approaches have been developed to utilize PINNs for applications in solid mechanics.
\cite{haghighat2020deep} trained PINN for small strain elasticity and elastoplasticity problems,
\cite{kadeethum2020physics} used the approach to solve the coupled Biot’s equations, and
\cite{rao2020physics} explored a mixed formulation for regular physics-informed neural networks with applications to small strain elastodynamics problems.
\cite{abueidda2020deep} utilized the residual formulation of PINNs to study  linear elasticity, hyperelasticity, and plasticity. 
Recently, the Deep Energy Method (DEM) was introduced, which defines the loss function not based on residual equations but based on energy minimization principles \citep{weinan2018deep, samaniego2020energy, nguyen2020deep}. Compared to classical PINN, this method has advantages because it only requires first order differentiation through the neural network. On the other hand it relies on accurate numerical integration techniques. With this technique \cite{nguyen2020deep} solve two and three dimensional finite-strain hyperelastic problems.
However as will be shown in this work, both the PINN and DEM struggle to resolve displacement and stress concentrations. In this context we propose the mixed Deep Energy Method (mDEM) which, by  requiring both displacements and stress outputs from the NN architecture, is able to resolve concentration features on comparable levels to FEM. Hence, the presented method aims to improve the forward capabilities of the physics-informed machine learning paradigm for solid mechanics applications as a first step towards efficient and accurate inverse problem calculations. Furthermore we provide some extension to the initial DEM formulation by offering a more reliable integration scheme based on Delaunay triangulation.

In section \ref{sec::1} the general finite-strain solid mechanics framework for hyperelasticity is introduced, and specialized to a Neo-Hookean type strain energy. A short introduction to neural networks, PINNs and DEMs specialized for 2D hyperelasticity is provided in Section \ref{sec::2}. The newly proposed mixed Deep Energy Method formulation as well as an accompanying integration scheme based on Delaunay triangulation are introduced in Section \ref{sec::3}. In Section \ref{sec::4} we test the approach on various computational experiments comparing against existing PINN- and DEM-based implementations. The paper is concluded in Section \ref{sec::conclusion}.

\section{Finite-strain hyperelasticity}\label{sec::1}

\begin{figure}
    \centering
    \includegraphics[scale=0.5]{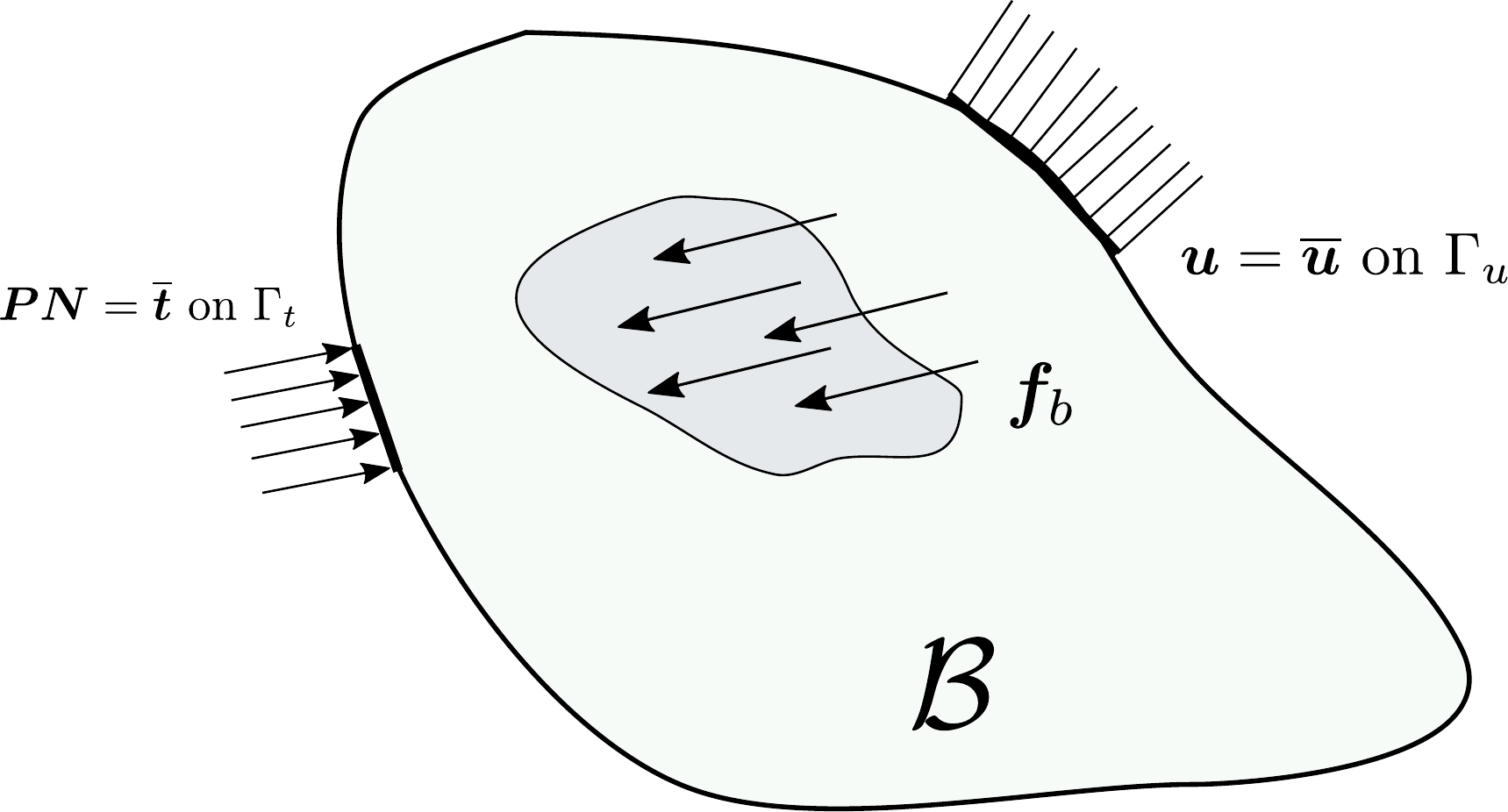}
    \caption{Solid domain with boundary conditions}
    \label{fig:pot}
\end{figure}

Let a bounded domain be occupied by an elastic body $\mathcal{B} \subset \mathbb{R}^{3}$. The body's boundary $\Gamma$ consists of two non-overlapping regions $\Gamma_{t}$ and $\Gamma_{u}$ such that $\Gamma = \Gamma_{t} \cup \Gamma_{u}$ (Fig. \ref{fig:pot}).
Let the referential and current position vectors of the body be denoted by $\bm{X}$ and $\bm{x}$ respectively.
The time-dependent motion between the two positions is then given by
\begin{equation}
\bm{x} = \bm{\varphi}(\bm{X},t) = \bm{X} + \bm{u}(\bm{X},t)
\end{equation} 
with $\bm{\varphi}(\bm{X},t)$ denoting the motion of the body and $\bm{u}$ the time-dependent displacement field. 
The gradient of the motion with respect to the initial position yield the deformation gradient
\begin{equation}
\bm{F} = \text{Grad} \bm{\varphi}(\bm{X}).
\end{equation}
The governing equation for quasi-static  problems in solid continua is the equilibrium equation, which in strong form reads
\begin{equation}\label{eq::Equilibrium}
\begin{aligned}
\text{Div} \bm{P} + \bm{f}_{b} &= \bm{0}, \quad \mathrm{in}\,\,\mathcal{B}
\end{aligned}
\end{equation}
with
\begin{equation}\label{eq::BoundaryCondition}
\begin{aligned}
\bm{u} &= \tilde{\bm{u}} \quad &&\text{on} \,\, \Gamma_{u} , \\
\bm{P} \cdot \bm{N} &= \tilde{\bm{t}} \quad &&\text{on} \,\, \Gamma_{t} ,
\end{aligned}
\end{equation}
where $\bm{P}$ and $\bm{f}_{b}$ denote the first Piola-Kirchhoff stress tensor and the body force vector respectively.
The body is in general subjected to two types of boundary conditions
with the outward normal $\bm{N}$, the prescribed displacement $\tilde{\bm{u}}$ on $\Gamma_{t}$ and the prescribed surface traction $\tilde{\bm{t}}$ on and $\Gamma_{u}$ as denoted in eq. \ref{eq::BoundaryCondition}.

In the hyperelastic framework, the existence of the strain energy function $\Psi$ is postulated, which enables the computation of the first Piola-Kirchhoff stress tensor by
\begin{equation}\label{eq::Constitutivelaw}
\bm{P} = \frac{\partial \Psi}{\partial \bm{F}}.
\end{equation}
In this paper we define a Neo-Hookean type strain energy which allows for compressibility as
 \begin{equation}
\Psi = \frac{1}{4} \lambda  ( \log J^{2} -1 - 2 \log J)  + \frac{1}{2} \mu (\text{tr}(\bm{C}) -  2 - 2 \log J),
\end{equation}
where $\lambda$ and $\mu$ are the Lam\'{e} constants and $J$ and $\bm{C}$ are given by $J = \det \bm{F}$ and $\bm{C} = \bm{F}^{T} \bm{F}$.
Considering only conservative loads and neglecting inertial contributions, the potential energy of the body is defined by
\begin{equation}\label{eq::Pot}
\Pi(\bm{\varphi}) = \int_{\mathcal{B}} \Psi \,dV - \int_{\mathcal{B}} \bm{f}_{b} \cdot \bm{\varphi} \,dV -
 \int_{\Gamma_{t}} \tilde{\bm{t}} \cdot \bm{\varphi} \,dA.
\end{equation}
The energy functional from eq. \ref{eq::Pot} can be minimized with regards to $\bm{\varphi}$
which yields the deformation that fulfills static equilibrium.

\section{Deep neural networks, Physics-informed Neural Networks  and the Deep Energy Method }\label{sec::2}

In this section we provide a brief overview of the neural networks formulation, Physics-informed Neural Networks and the Deep Energy Method.
Neural networks are in general composed of one input, one output and $n_{D}-1$ hidden layers.
Let the weights and biases of the $k^{\text{th}}$ layer be denoted by $\bm{W}^{k} $ and $\bm{b}^{k}$. 
Consider that the $k^{\text{th}}$ hidden layer transfers some output $\bm{x}^{k}$ to the $(k+1)^{\text{th}}$ layers which applies an affine transformation
\begin{equation}\label{eq::NNO}
    \mathcal{L} (\bm{x}^{k})  = \bm{W}^{k+1} \bm{x}^{k} + \bm{b}^{k+1}.
\end{equation}
and some activation activation function $\sigma$ to it. We consider the activation function of the last layer to be identity.
Since, equation (\ref{eq::NNO}) is applied in every layer of the network an input $\bm{x}$ yields a network output of the form
\begin{equation}
    \hat{\bm{y}}(\bm{x}) = (\mathcal{L}_{k} \circ \sigma \circ \mathcal{L}_{k-1}\circ \cdots \circ \sigma \circ \mathcal{L}_{1})(\bm{x})
\end{equation}
where $\circ$ is a composition operator. 
The goal of neural networks is to find the optimal value set of trainable parameters $\bm{\Theta} = \lbrace \bm{W}^{k}, \bm{b}^{k}  \rbrace_{k=1}^{n_{D}}$ such that the networks provides the best fit for the input-output mapping.
This is achieved 
by following an optimization procedure defined over some loss function $J(\bm{\Theta} )$
\begin{equation}\label{eq::0_optimization}
    \bm{\Theta}^{\star} = \argmin_{\bm{\Theta}} J(\bm{\Theta} ).
\end{equation}
This problem is typically solved in an iterative manner by employing a stochastic gradient-descent approach. A review on neural network optimization methods is provided in \cite{bottou2018optimization}.
For more information on neural networks we refer to \cite{goodfellow2016deep}. \\
PINNs define a global shape function for the field(s) of interest over the computational domain and then require the stresses derived with automatic differentiation to fulfill a residual based on the strong form, including the equilibrium equation (\ref{eq::Equilibrium}) as well as the corresponding boundary conditions (\ref{eq::BoundaryCondition}). The computational domain is spanned by a set of collocation points. Here we will specialize the overview of PINN in the context of hyperelasticity. In this case the unkown field that will be approximated through PINN is the displacement field.
In absence of body forces, the residual is of the form
\begin{equation}
    \bm{\mathcal{R}} = \text{Div} \bm{P} = \bm{0}. 
\end{equation}

The collocation points hereby act as inputs to the neural network which yield the outputs $\bm{z}(\bm{X}, \bm{\Theta})$. The outputs $\bm{z}(\bm{X}, \bm{\Theta})$ can either be subjected to 
\begin{equation}\label{eq::PINNOutput1}
\hat{\bm{u}}(\bm{X}, \bm{\Theta}) = \bm{A}(\bm{X}) + \bm{B}(\bm{X}) \circ \bm{z}(\bm{X}, \bm{\Theta})
\end{equation}
where $\bm{A}$ and $\bm{B}$ are a-priori chosen in a way such that the displacement prediction $\hat{\bm{u}}$ fulfills the displacement boundary conditions acting on the body. Here, $\bm{A}$ and $\bm{B}$ can either chosen analytically or can be neural networks themselves. For a detailed discussion refer to \cite{rao2020physics}. Alternately, we can choose
\begin{equation}\label{eq::PINNOutput2}
\hat{\bm{u}}(\bm{X}, \bm{\Theta}) = \bm{z}(\bm{X}, \bm{\Theta})
\end{equation}
where the boundary conditions need to be trained. From the displacements the deformation gradient can be obtained by
\begin{equation}\label{eq::DeformationGradientPINN}
\begin{aligned}
\hat{\bm{F}}(\bm{X}, \bm{\Theta}) &= \bm{I} + \text{Grad}\hat{\bm{u}}( \bm{X}, \bm{\Theta}).
\end{aligned}
\end{equation}
Consequently, the stresses $\bm{P}(\hat{\bm{F}}(\bm{X}, \bm{\Theta}))$  can be computed following their definitions eq. \ref{eq::Constitutivelaw}.

Consider a set of domain training points $\lbrace \bm{X}_{\mathcal{R}}^{i} \rbrace_{i=1}^{N_{\mathcal{R}}}$, displacement boundary training points $\lbrace \bm{X}_{u}^{i} \rbrace_{i=1}^{N_{u}}$ and traction boundary training points $\lbrace \bm{X}_{t}^{i} \rbrace_{i=1}^{N_{t}}$  where $N_{\mathcal{R}}$, $N_{u}$ and $N_{t}$ denote the number of points, respectively. The PINN loss function is then defined as
\begin{equation}\label{eq::PINN_optimization}
\begin{aligned}
        \bm{\Theta}^{\star} &= \argmin_{\bm{\Theta}}   \qquad W_{\mathcal{R}} \text{MSE}_{\mathcal{R}} \left(\lbrace \bm{X}_{\mathcal{R}}^{i} \rbrace_{i=1}^{N_{\mathcal{R}}}, \bm{\Theta}  \right) + W_{t} \text{MSE}_{t} \left(\lbrace \bm{X}_{t}^{i} \rbrace_{i=1}^{N_{t}}, \bm{\Theta}  \right) + \underbrace{W_{u} \text{MSE}_{u} \left(\lbrace \bm{X}_{u}^{i} \rbrace_{i=1}^{N_{u}}, \bm{\Theta}  \right)}_{ \text{If not a-priori fulfilled}} 
\end{aligned}
\end{equation}
with some weight values $W_{\mathcal{R}}$, $W_{t}$ and $W_{u}$ and specializing to a 2D case
\begin{equation}
\begin{aligned}
\text{MSE}_{\mathcal{R}} \left(\lbrace \bm{X}_{\mathcal{R}}^{i} \rbrace_{i=1}^{N_{\mathcal{R}}}, \bm{\Theta}  \right) &=\frac{1}{N_{\mathcal{R}}} \sum_{i=1}^{N_{\mathcal{R}}} \abs{\frac{\partial}{\partial X_{1}} P_{11}(\hat{\bm{F}}(\bm{X}_{\mathcal{R}}^{i}, \bm{\Theta})) +  \frac{\partial}{\partial X_{2}} P_{12}(\hat{\bm{F}}(\bm{X}_{\mathcal{R}}^{i}, \bm{\Theta})) }^{2} \\
&+ \frac{1}{N_{\mathcal{R}}} \sum_{i=1}^{N_{\mathcal{R}}} \abs{ \frac{\partial}{\partial X_{1}} P_{21}(\hat{\bm{F}}(\bm{X}_{\mathcal{R}}^{i}, \bm{\Theta})) +  \frac{\partial}{\partial X_{2}} {P}_{22}(\hat{\bm{F}}(\bm{X}_{\mathcal{R}}^{i}, \bm{\Theta}))  }^{2},
 \\
    \text{MSE}_{t} \left(\lbrace \bm{X}_{t}^{i} \rbrace_{i=1}^{N_{t}}, \bm{\Theta}  \right) &= \frac{1}{N_{t}} \sum_{i=1}^{N_{t}} \abs{ \bm{P}(\hat{\bm{F}}(\bm{X}_{t}^{i}, \bm{\Theta})) \bm{N}_{i} - \overline{\bm{t}}(\bm{X}_{t}^{i})}^{2},
 \\
    \text{MSE}_{u} \left(\lbrace \bm{X}_{u}^{i} \rbrace_{i=1}^{N_{u}}, \bm{\Theta}  \right) &= \frac{1}{N_{u}} \sum_{i=1}^{N_{u}} \abs{ \hat{\bm{u}}(\bm{X}_{u}^{i}, \bm{\Theta}) - \bm{u}(\bm{X}_{u}^{i})}^{2}.
\end{aligned}    
\end{equation}

\begin{figure}[hbtp]
\centering
\includegraphics[scale=0.8]{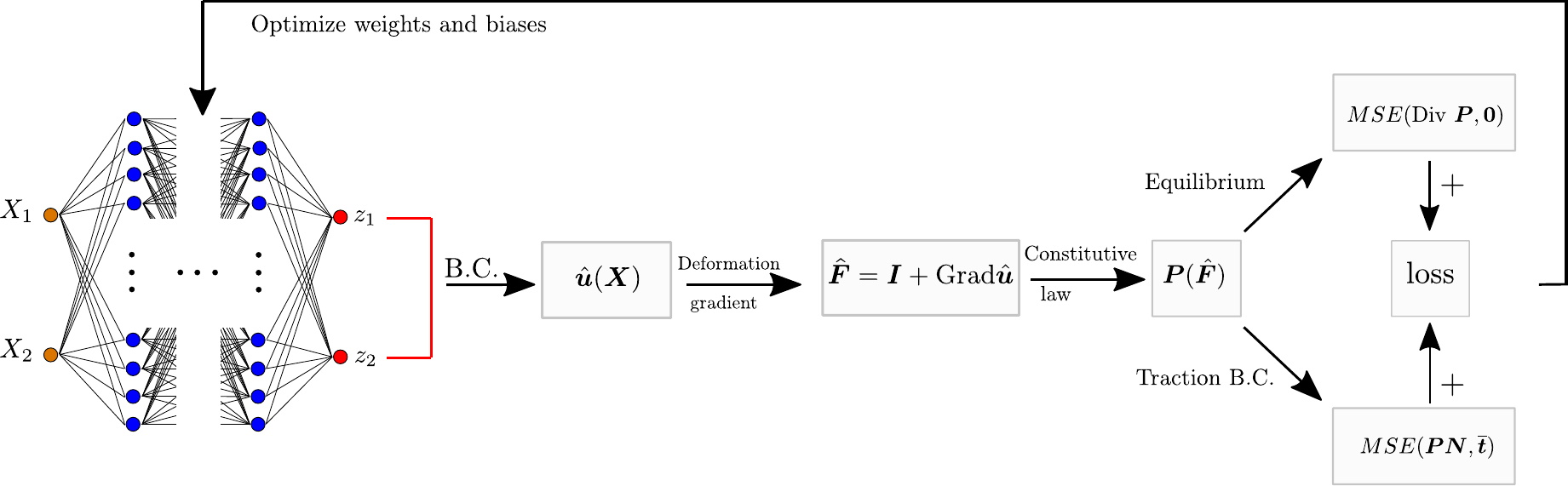}
\caption{PINN overview}\label{fig::PINN}
\end{figure}

The PINN concept is summarized in Figure \ref{fig::PINN}.
Maintaining some of the architecture of PINNs,
\cite{nguyen2020deep} proposed an alternative approach. The neural network formulation is still used to solve quasi-static equilibrium problems by utilizing the network as a global shape function for the displacement over the body.
However instead of defining the residual of the governing PDE as the loss function, the authors define the loss as the potential energy of the solid body, as defined in equation \ref{eq::Pot}. By following that, the minimization of the loss function over some collocation points that span the computational domain and its boundary, through stationarity of the potential energy in nonlinear problems, leads to an approximate solution.    The advantage of this approach is that only first order automatic differentiation is needed for the solution procedure, instead of second order as required in PINN. This, can potentially lead to quicker convergence and higher accuracy. As a downside the method requires an effective integration procedure over the domain spanned by the collocation points.
In essence DEM uses the same concept as PINNs for the input and the output to the neural network, i.e. collocation point positions as input and the output is defined by either eq. \ref{eq::PINNOutput1} or \ref{eq::PINNOutput2}.
In addition to the deformation gradient (eq. \ref{eq::DeformationGradientPINN}), DEM requires a formulation for the motion which is given by
\begin{equation}
\begin{aligned}
\hat{\bm{\varphi}}(\bm{X},\bm{\Theta}) &= \bm{X} + \hat{\bm{u}}( \bm{X}, \bm{\Theta}).
\end{aligned}
\end{equation}
Consider a set of domain training points $\lbrace \bm{X}_{\Pi}^{i} \rbrace_{i=1}^{N_{\Pi}}$ and boundary training points $\lbrace \bm{X}_{u}^{i} \rbrace_{i=1}^{N_{u}}$ where $N_{\Pi}$ and $N_{u}$ denote the number of points, respectively.
Hence, the loss function can be written as
\begin{equation}\label{eq::DEM_optimization}
\begin{aligned}
        \bm{\Theta}^{\star} &= \argmin_{\bm{\Theta}}   \qquad  &&\Pi (\hat{\bm{\varphi}}(\lbrace \bm{X}_{\Pi}^{i} \rbrace_{i=1}^{N_{\Pi}},\bm{\Theta})) +  \underbrace{W_{u} \text{MSE}_{u} \left(\lbrace \bm{X}_{u}^{i} \rbrace_{i=1}^{N_{u}}, \bm{\Theta}  \right)}_{ \text{If not a-priori fulfilled}} 
\end{aligned}
\end{equation}
with a weight value $W_{u}$ and where 
\begin{equation}
\begin{aligned}\Pi (\hat{\bm{\varphi}}(\lbrace \bm{X}_{\Pi}^{i} \rbrace_{i=1}^{N_{\Pi}},\bm{\Theta})) &=
\int_{\mathcal{B}}   \Psi (\hat{\bm{F}}(\lbrace \bm{X}_{\Pi}^{i} \rbrace_{i=1}^{N_{\Pi}},\bm{\Theta})) \,dV - \int_{\mathcal{B}} \bm{f}_{b} \cdot \hat{\bm{\varphi}}(\lbrace \bm{X}_{\Pi}^{i} \rbrace_{i=1}^{N_{\Pi}},\bm{\Theta}) \,dV -
 \int_{\Gamma_{t}} \tilde{\bm{t}} \cdot \hat{\bm{\varphi}}(\lbrace \bm{X}_{\Pi}^{i} \rbrace_{i=1}^{N_{\Pi}},\bm{\Theta}) \,dA \\
    \text{MSE}_{u} \left(\lbrace \bm{X}_{u}^{i} \rbrace_{i=1}^{N_{u}}, \bm{\Theta}  \right) &= \frac{1}{N_{u}} \sum_{i=1}^{N_{u}} \abs{ \hat{\bm{u}}(\bm{X}_{u}^{i}, \bm{\Theta}) - \bm{u}(\bm{X}_{u}^{i})}^{2}.
\end{aligned}    
\end{equation}

\begin{figure}[hbtp]
\centering
\includegraphics[scale=0.8]{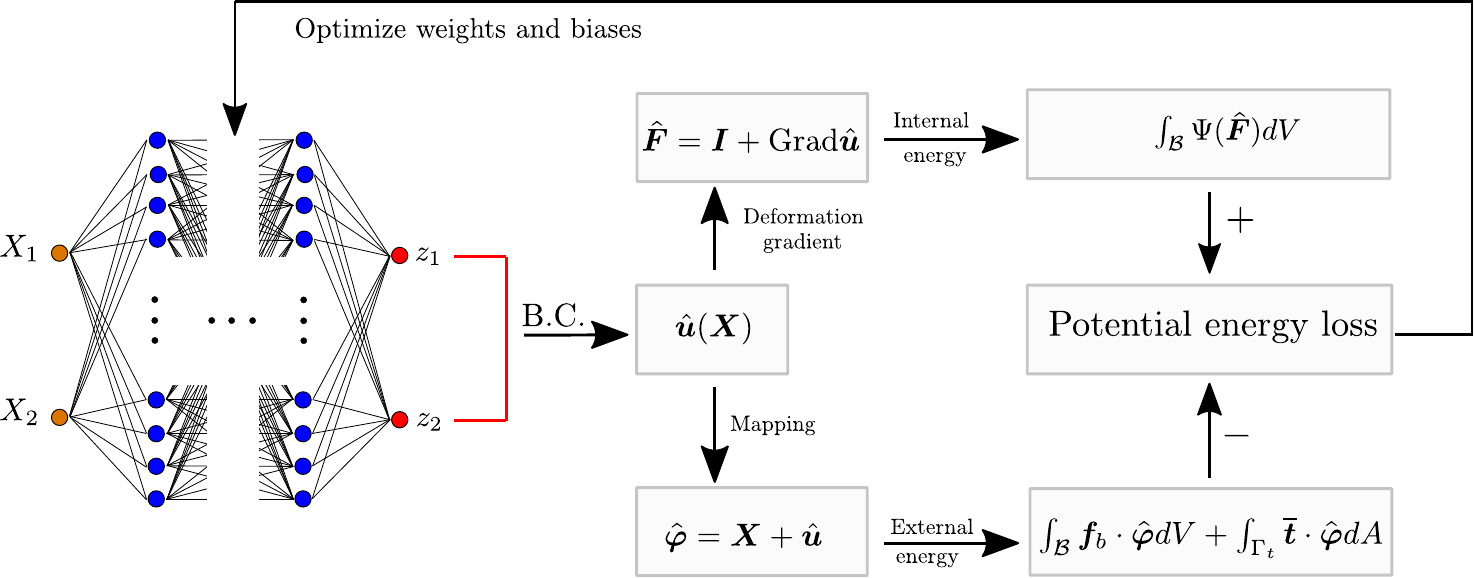}
\caption{DEM overview}\label{fig::DEM}
\end{figure}
The process is summarized in Figure \ref{fig::DEM}.

\section{Mixed Deep Energy Method (MDEM)}\label{sec::3}
For solid mechanics and multiphysics applications, problematic for the DEM and PINN formulations is that they average out stress and displacement concentrations as will be shown in Section \ref{sec::4}. In order to circumvent these issues we propose the Mixed Deep Energy Method (mDEM) which introduces stresses as additional collocation point outputs, which in turn have to fulfill the traction boundary conditions. 
Furthermore to reduce the computation time, training a different neural network for each individual output, as e.g. proposed in \cite{haghighat2020deep}, has been explicitly avoided in this work.
Hence, in a 2D setting, in addition to the two outputs $\bm{z}$ from DEM we also obtain four more outputs $\bm{Z}$ corresponding to the first Piola-Kirchhoff stress components. Equivalently these outputs can individually be trained to directly fulfill the corresponding Dirichlet and Neumann boundary conditions respectively, or can be subjected to 
\begin{equation}
\begin{aligned}
    \hat{\bm{u}}(\bm{X}, \bm{\Theta}) &= \bm{A}(\bm{X}) + \bm{B}(\bm{X}) \circ \bm{z}(\bm{X}, \bm{\Theta}) \\
    \hat{\bm{P}}(\bm{X}, \bm{\Theta}) &= \bm{C}(\bm{X}) + \bm{D}(\bm{X}) \circ \bm{Z}(\bm{X}, \bm{\Theta})
\end{aligned}
\end{equation}
such that these conditions are fulfilled a-priori.
In order to connect the displacements and stresses, mDEM requires training of the constitutive behavior by enforcing
\begin{equation}
    \hat{\bm{P}}(\bm{X}, \bm{\Theta}) = \bm{P}(\hat{\bm{F}}(\bm{X}, \bm{\Theta}))
\end{equation}
where the latter term is derived from the displacement output and the hyperleastic constitutive behavior of equation (\ref{eq::Constitutivelaw}).

\begin{figure}[hbtp]
\centering
\includegraphics[scale=0.8]{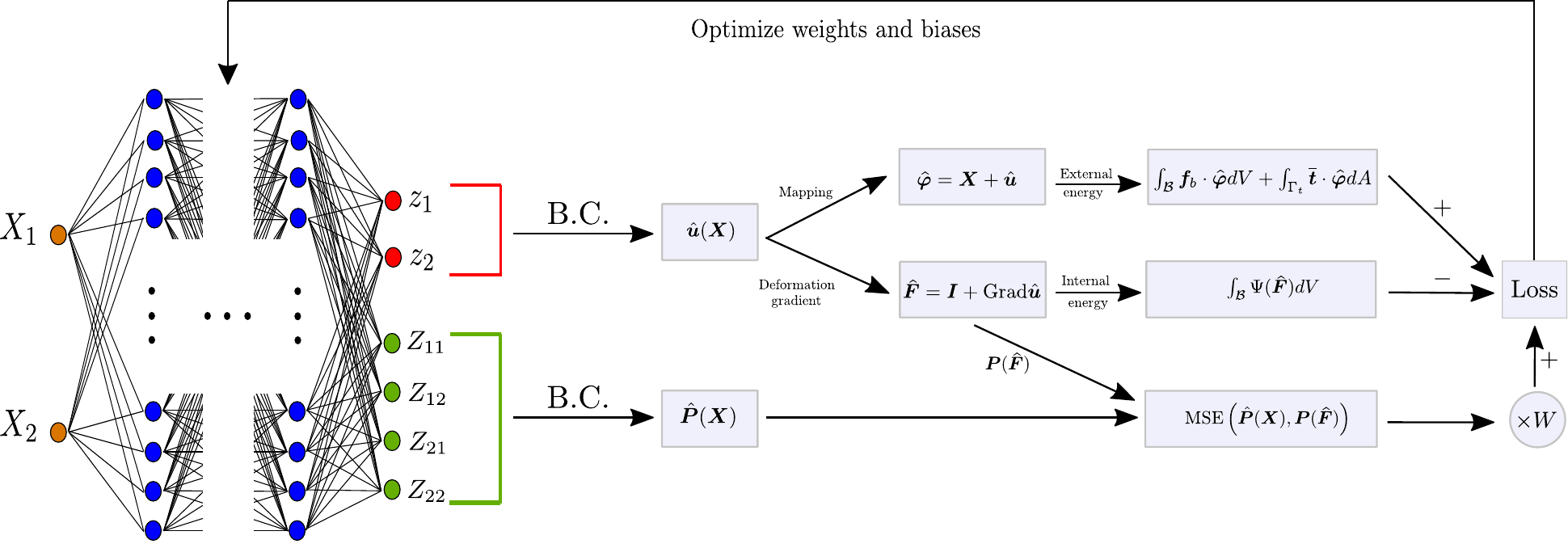}
\caption{DEM overview}\label{fig::mDEM}
\end{figure}

Therefore we consider a set of domain training points $\lbrace \bm{X}_{\Pi}^{i} \rbrace_{i=1}^{N_{\Pi}}$, displacement boundary condition points $\lbrace \bm{X}_{u}^{i} \rbrace_{i=1}^{N_{u}}$, and traction boundary condition points $\lbrace \bm{X}_{t}^{i} \rbrace_{i=1}^{N_{t}}$, where $N_{\Pi}$, $N_{u}$ and $N_{t}$ denote the number of points.
The loss function can the be written as
\begin{equation}\label{eq::mDEM_optimization}
\begin{aligned}
        \bm{\Theta}^{\star} &= \argmin_{\bm{\Theta}}   \qquad  &&\Pi (\hat{\bm{\varphi}}(\lbrace \bm{X}_{\Pi}^{i} \rbrace_{i=1}^{N_{\Pi}},\bm{\Theta})) + W_{P} \text{MSE}_{P} \left(\lbrace \bm{X}_{\Pi}^{i} \rbrace_{i=1}^{N_{\Pi}}, \bm{\Theta}  \right) \\& &&+
        \underbrace{W_{u} \text{MSE}_{u} \left(\lbrace \bm{X}_{u}^{i} \rbrace_{i=1}^{N_{u}}, \bm{\Theta}  \right)}_{ \text{If not a-priori fulfilled}} +
        \underbrace{W_{t} \text{MSE}_{t} \left(\lbrace \bm{X}_{t}^{i} \rbrace_{i=1}^{N_{t}}, \bm{\Theta}  \right)}_{ \text{If not a-priori fulfilled}} 
\end{aligned}
\end{equation}
with weight values $W_{u}$,$W_{t}$, $W_{P}$ and where 
\begin{equation}
\begin{aligned}\Pi (\hat{\bm{\varphi}}(\lbrace \bm{X}_{\Pi}^{i} \rbrace_{i=1}^{N_{\Pi}},\bm{\Theta})) &=
\int_{\mathcal{B}}   \Psi (\hat{\bm{F}}(\lbrace \bm{X}_{\Pi}^{i} \rbrace_{i=1}^{N_{\Pi}},\bm{\Theta})) \,dV - \int_{\mathcal{B}} \bm{f}_{b} \cdot \hat{\bm{\varphi}}(\lbrace \bm{X}_{\Pi}^{i} \rbrace_{i=1}^{N_{\Pi}},\bm{\Theta}) \,dV -
 \int_{\Gamma_{t}} \tilde{\bm{t}} \cdot \hat{\bm{\varphi}}(\lbrace \bm{X}_{\Pi}^{i} \rbrace_{i=1}^{N_{\Pi}},\bm{\Theta}) \,dA, \\
     \text{MSE}_{P} \left(\lbrace \bm{X}_{\Pi}^{i} \rbrace_{i=1}^{N_{\Pi}}, \bm{\Theta}  \right) &= \frac{1}{N_{\Pi}} \sum_{i=1}^{N_{\Pi}} \abs{ \hat{\bm{P}}(\bm{X}_{\Pi}^{i}, \bm{\Theta})  - \bm{P}(\hat{\bm{F}}(\bm{X}_{\Pi}^{i}, \bm{\Theta}))}^{2}, \\
    \text{MSE}_{u} \left(\lbrace \bm{X}_{u}^{i} \rbrace_{i=1}^{N_{u}}, \bm{\Theta}  \right) &= \frac{1}{N_{u}} \sum_{i=1}^{N_{u}} \abs{ \hat{\bm{u}}(\bm{X}_{u}^{i}, \bm{\Theta}) - \bm{u}(\bm{X}_{u}^{i})}^{2}, \\
    \text{MSE}_{t} \left(\lbrace \bm{X}_{t}^{i} \rbrace_{i=1}^{N_{t}}, \bm{\Theta}  \right) &= \frac{1}{N_{t}} \sum_{i=1}^{N_{t}} \abs{ \hat{\bm{P}}(\bm{X}_{t}^{i}, \bm{\Theta}) \bm{N}_{i} - \overline{\bm{t}}(\bm{X}_{t}^{i})}^{2}.
\end{aligned}    
\end{equation}
The process is visualized in Figure \ref{fig::mDEM}.

\subsection{Delauney integration}
\begin{figure}
\begin{subfigure}[b]{0.5\linewidth}
\centering
\includegraphics[scale=0.2]{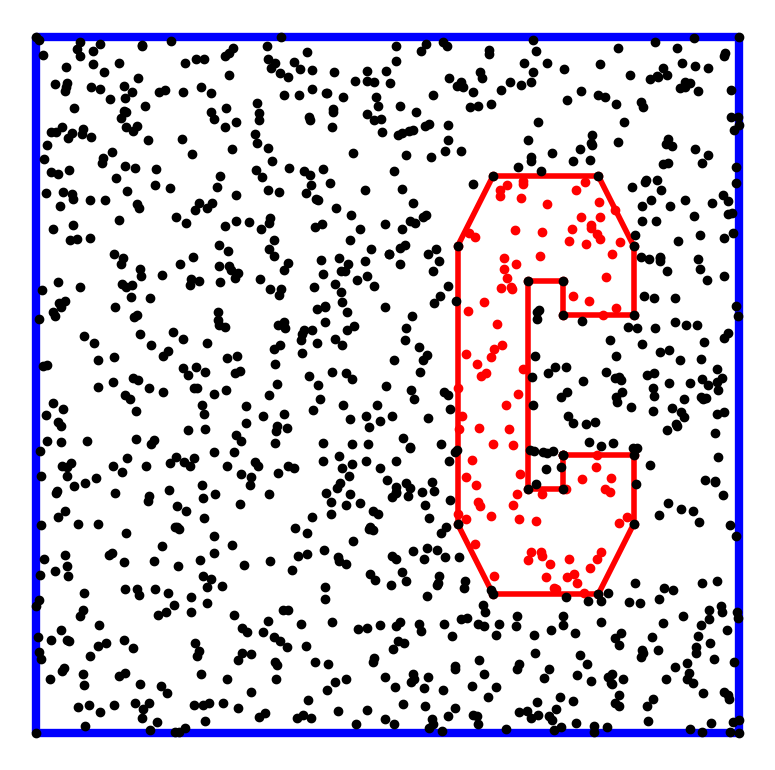} 
\caption{a}
\end{subfigure}%
\begin{subfigure}[b]{.5\linewidth}
\centering
\includegraphics[scale=0.2]{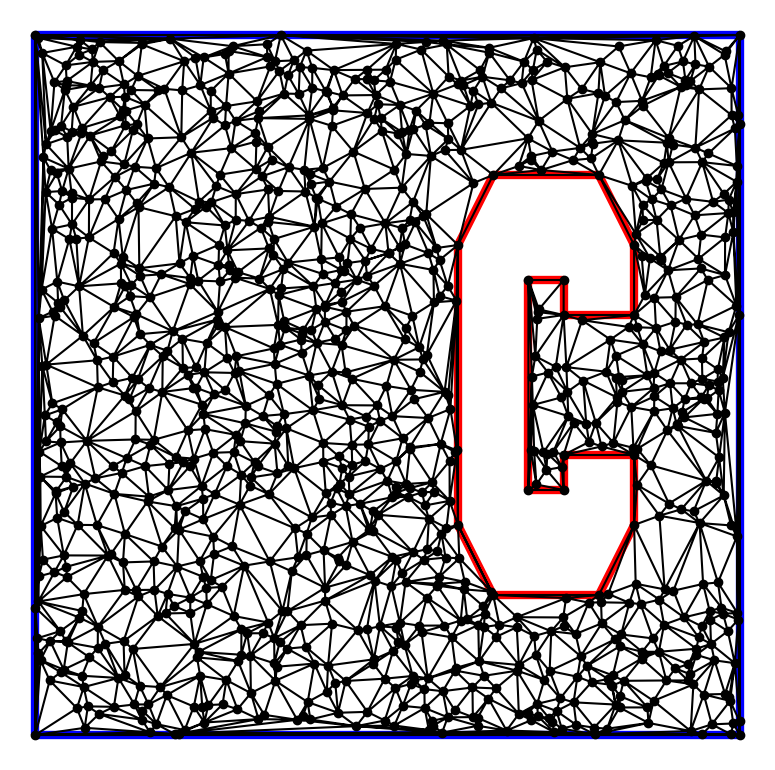} 
\caption{b}
\end{subfigure}
\caption{Constraint Voronoi tesselation for integration of defected area.}\label{fig:ConstraintVoronoi}
\end{figure}

In the DEM paper \citep{nguyen2020deep}, the authors explore different numerical integration techniques for the evaluation of the integral including Trapezoidal and Simpson's rule as well as a Monte Carlo integration version. Even though the presented techniques prove to be sufficient for the investigated numerical examples, they are difficult to employ when the sample points need to represent solid continua with fine geometric features, which commonly lead to stress concentrations. Hence, in this paper we employ a form of Delaunay-integration to resolve this issue. The computational domain is tesselated based on its respective sample positions resulting in $n_{e}$ triangles or tetrahedra. 
Given a triangle $j$ in two dimensions, with the three points denoted by $P_{i} = {x_{i}, y_{i}}$ and a corresponding function value $f_{i}$ with $i=1,2,3$ for the field of interest given at each point. Then define $J_{j}$ as the area of the triangle and let
\begin{equation}
\overline{f_{\bigtriangleup_{j}}} = \frac{f_{1}+f_{2}+f_{3}}{3}
\end{equation}
 denote the mean function value of an element.
Following eq. \ref{eq::Pot}, the potential energy of the domain is then approximated by
\begin{equation}
\begin{aligned}
\Pi(\bm{\varphi}) &= \int_{\mathcal{B}} \Psi \,dV - \int_{\mathcal{B}} \bm{f}_{b} \cdot \bm{\varphi} \,dV -
 \int_{\Gamma_{t}} \tilde{\bm{t}} \cdot \bm{\varphi} \,dA \\
&= \sum_{i=1}^{n_{e}} \int_{\bigtriangleup_{i}} \Psi_{\bigtriangleup_{i}} d \bigtriangleup_{i} - \sum_{i=1}^{n_{e}} \int_{\bigtriangleup_{i}} \left[ \bm{f}_{b} \cdot \bm{\varphi}\right]_{\bigtriangleup_{i}} d \bigtriangleup_{i} -
 \int_{\Gamma_{t}} \tilde{\bm{t}} \cdot \bm{\varphi} \,dA
 \\
 &=  \sum_{i=1}^{n_{e}}  J_{i}  \overline{\Psi_{\bigtriangleup_{i}}} - \sum_{i=1}^{n_{e}} J_{i}  \overline{\left[ \bm{f}_{b} \cdot \bm{\varphi}\right]_{\bigtriangleup_{i}}}  - \sum_{i=1}^{N_{b}}w_{i} \left[ \tilde{\bm{t}}_{i} \cdot \bm{\varphi}_{i} \right]
\end{aligned}
\end{equation}
where $w_{i}$ represents some weights to a common numerical integration technique such as Simpson's rule.
An example of a constraint Voronoi mesh for a  plate with a "C"-shaped hole (red zone) is displayed in Figure \ref{fig:ConstraintVoronoi}.

\section{Numerical examples}\label{sec::4}
In this section we compare the solutions of mDEM to DEM and PINN as well as to a Finite-Element Method (FEM) solution which will be taken as ground truth. The FEM results were obtained using the Fenics framework \citep{AlnaesBlechta2015a}.
The deep learning formulations were implemented\footnote{Codes will be made public after acceptance of this paper.} in 
Pytorch \citep{NEURIPS2019_9015} and the network parameters were optimized using the
Adam optimizer \citep{kingma2014adam} and the limited Broyden–Fletcher–Goldfarb–Shanno (LFBGS) algorithm consecutively. In the following, we use analytical formulations to fulfill the displacement boundary conditions of the applications a-priori, while the traction boundaries are learned from data. All neural networks are composed of 6 hidden layers with 60 neurons each. Furthermore all networks are trained with an Adam and LFBGS learning rate of $1e-3$.

\subsection{Uniaxial loading}\label{sec::UniaxialTension}
\begin{figure}[hbtp]
\centering
\begin{subfigure}[b]{0.5\linewidth}
\centering
\includegraphics[scale=0.35]{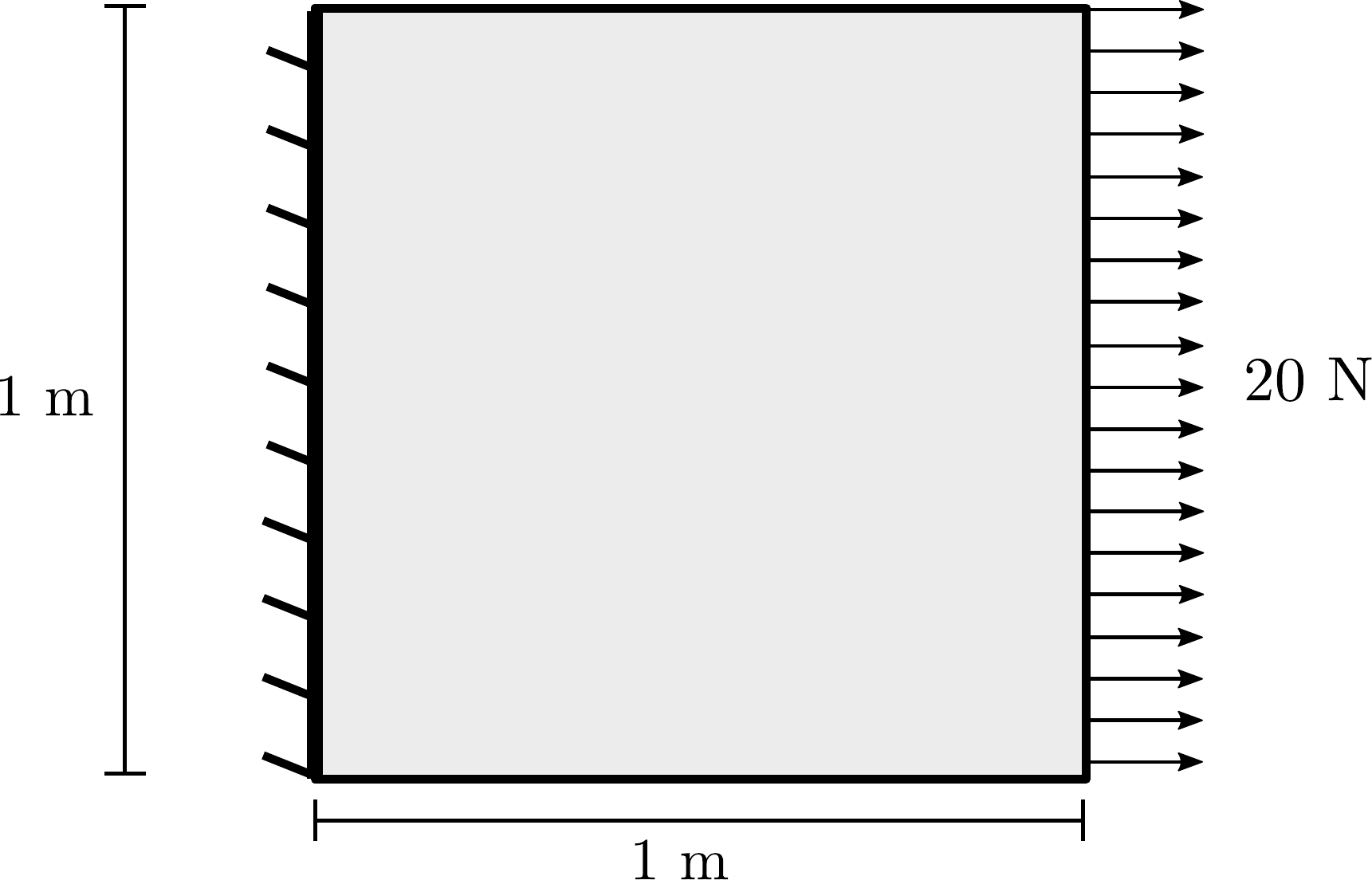}
\caption{}\label{fig::UniProbImage}
\end{subfigure}%
\begin{subfigure}[b]{0.5\linewidth}
\centering
\includegraphics[scale=0.23]{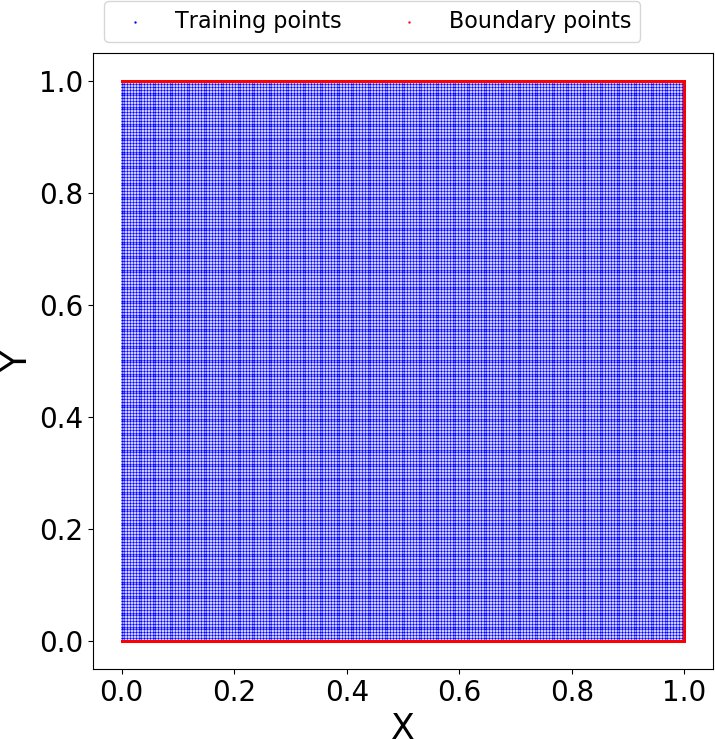}
\caption{}\label{fig::UniTrainPoints}
\end{subfigure}%
\caption{Uniaxial loading problem and training point positions.}\label{fig::UniaxialTension}
\end{figure}

\begin{figure}[b!]
\begin{subfigure}[b]{0.5\linewidth}
\centering
\includegraphics[scale=0.19]{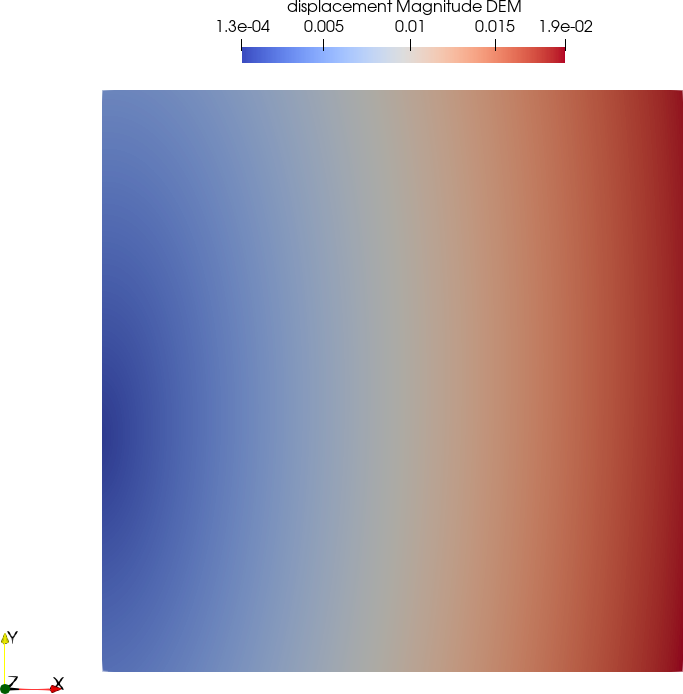} 
\caption{DEM displacement magnitude}
\end{subfigure}%
\begin{subfigure}[b]{.5\linewidth}
\centering
\includegraphics[scale=0.19]{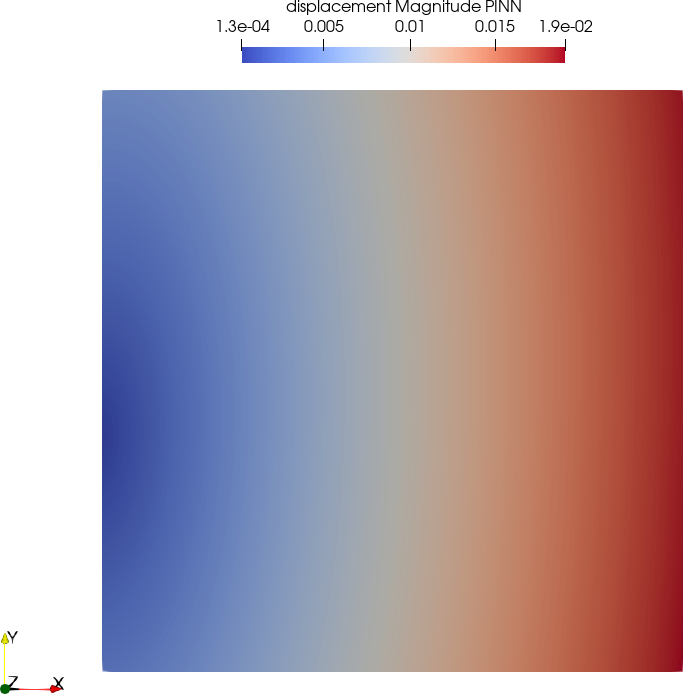} 
\caption{PINN displacement magnitude}
\end{subfigure}

\begin{subfigure}[b]{0.5\linewidth}
\centering
\includegraphics[scale=0.19]{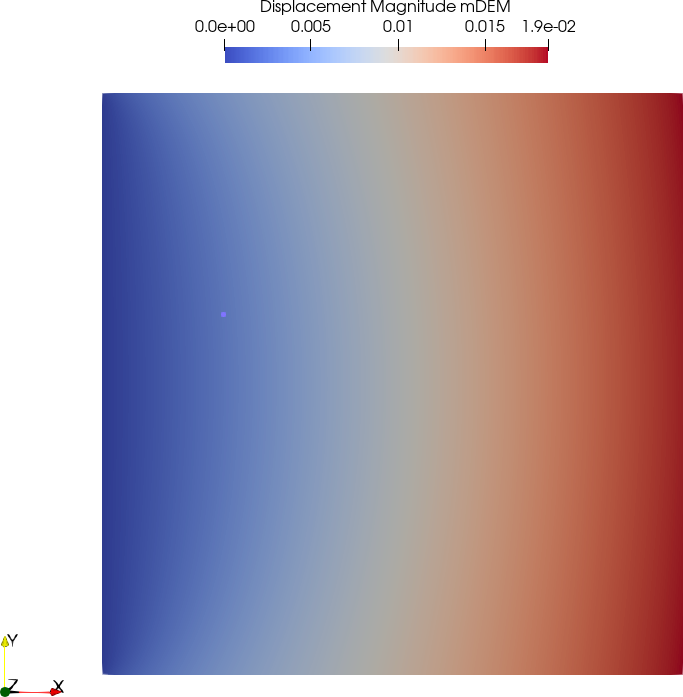} 
\caption{mDEM displacement magnitude}
\end{subfigure}%
\begin{subfigure}[b]{0.5\linewidth}
\centering
\includegraphics[scale=0.19]{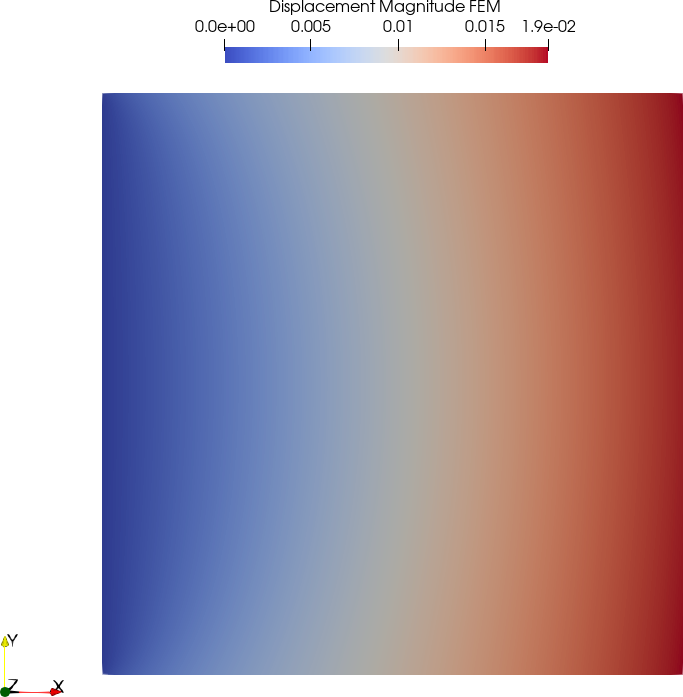} 
\caption{FEM displacement magnitude}
\end{subfigure}
\caption{Uniaxial loading problem displacement magnitude.}\label{fig:Prob1DispMagnitude}
\end{figure}

\begin{figure}
\begin{subfigure}[b]{0.5\linewidth}
\centering
\includegraphics[scale=0.2]{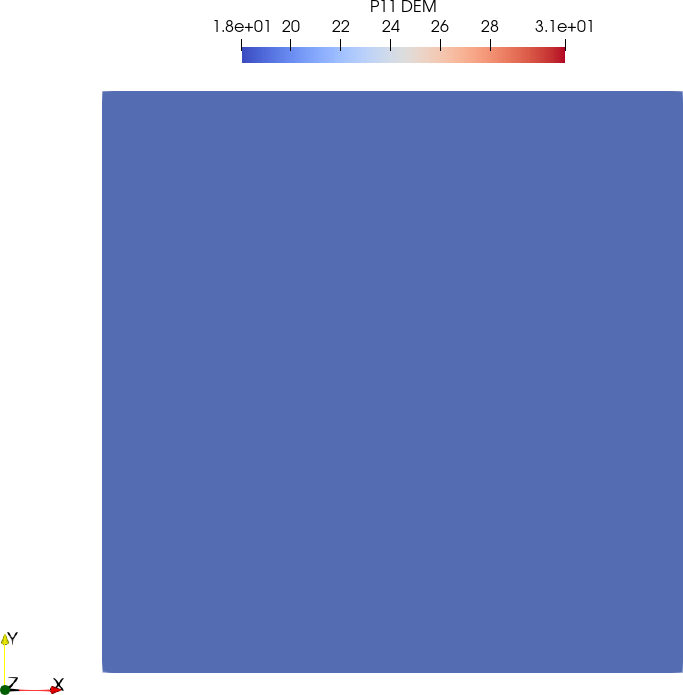} 
\caption{DEM $P_{11}$}
\end{subfigure}%
\begin{subfigure}[b]{.5\linewidth}
\centering
\includegraphics[scale=0.2]{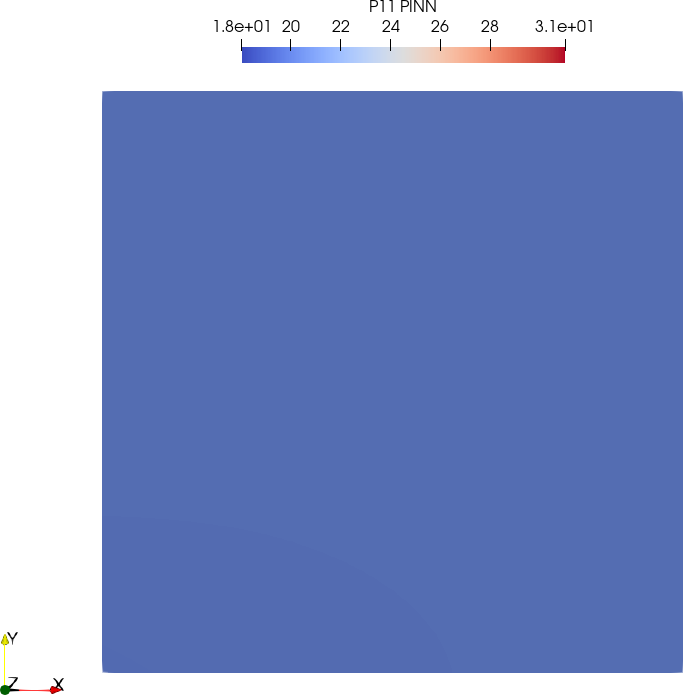} 
\caption{PINN $P_{11}$}
\end{subfigure}

\begin{subfigure}[b]{0.5\linewidth}
\centering
\includegraphics[scale=0.2]{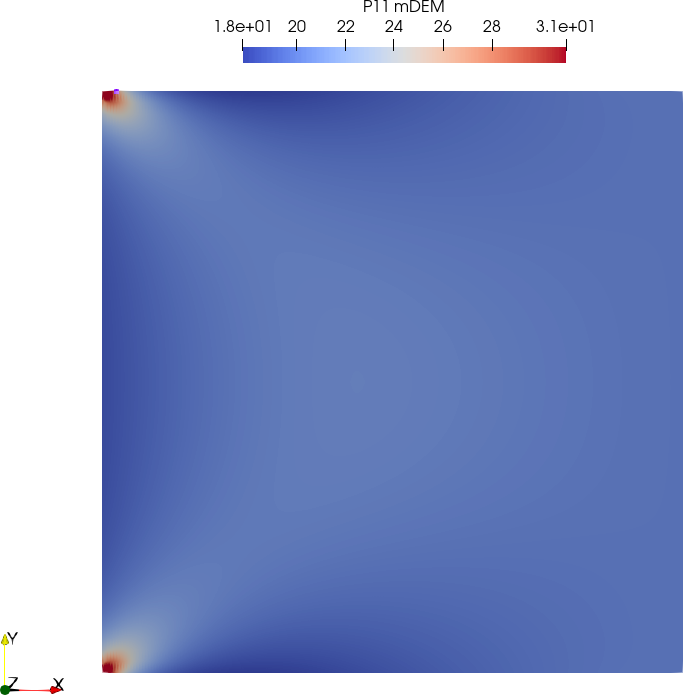} 
\caption{mDEM $P_{11}$}
\end{subfigure}%
\begin{subfigure}[b]{0.5\linewidth}
\centering
\includegraphics[scale=0.2]{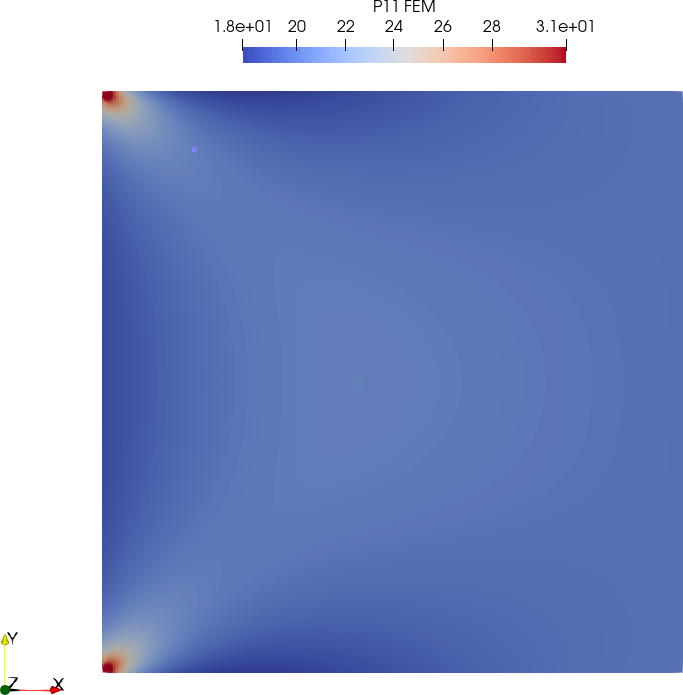} 
\caption{FEM $P_{11}$}
\end{subfigure}

\begin{subfigure}[b]{0.5\linewidth}
\centering
\includegraphics[scale=0.2]{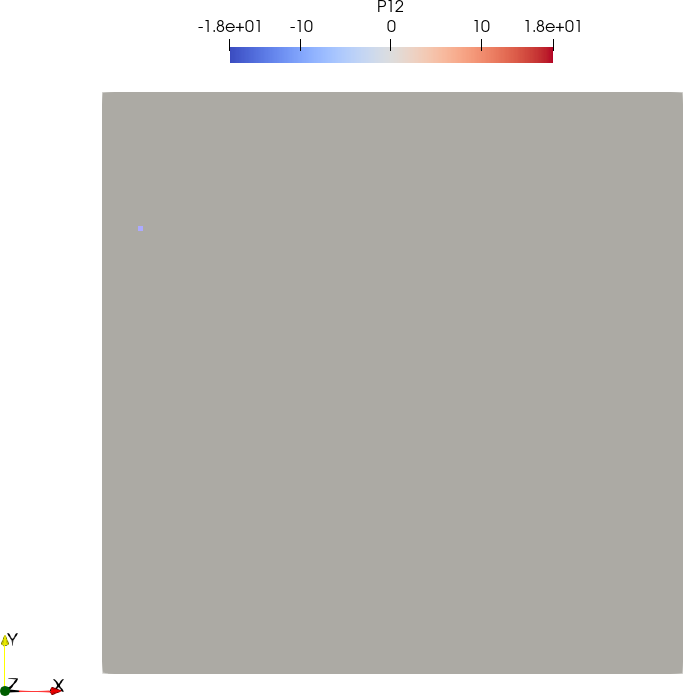} 
\caption{DEM $P_{12}$}
\end{subfigure}%
\begin{subfigure}[b]{.5\linewidth}
\centering
\includegraphics[scale=0.2]{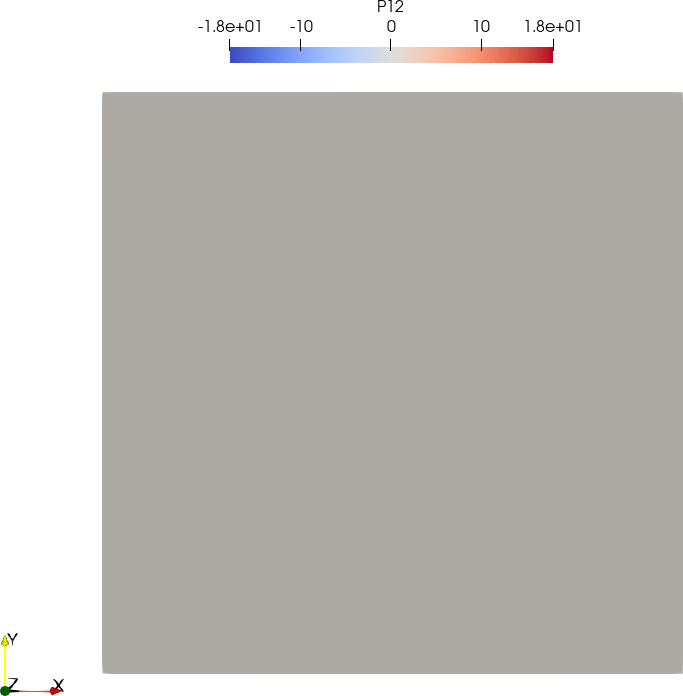} 
\caption{PINN $P_{12}$}
\end{subfigure}

\begin{subfigure}[b]{0.5\linewidth}
\centering
\includegraphics[scale=0.2]{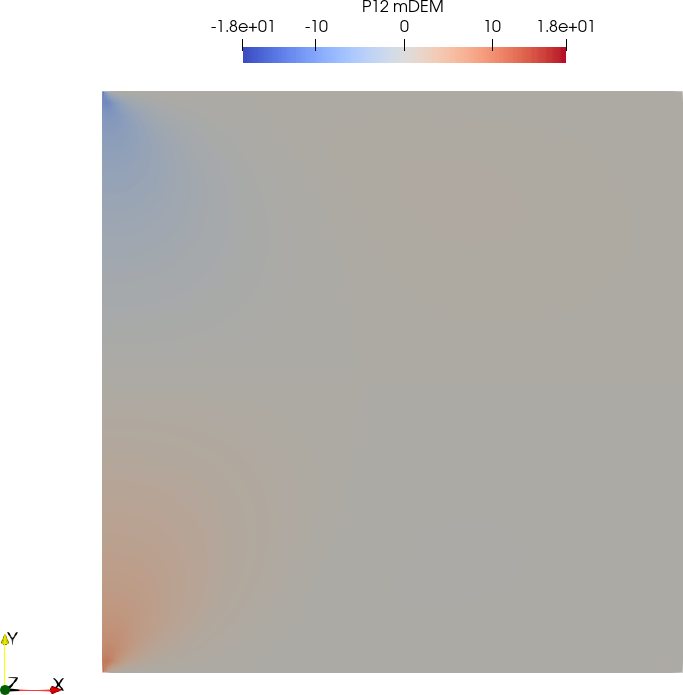} 
\caption{mDEM $P_{12}$}
\end{subfigure}%
\begin{subfigure}[b]{0.5\linewidth}
\centering
\includegraphics[scale=0.2]{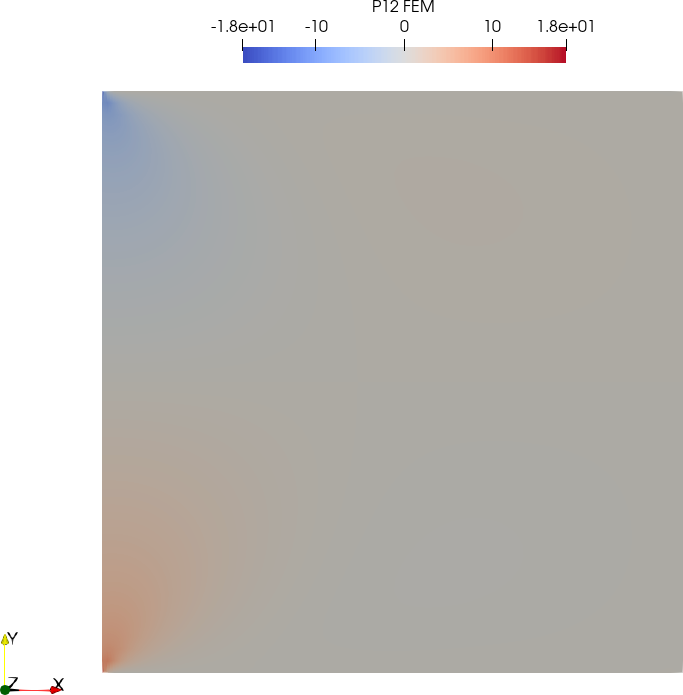} 
\caption{FEM $P_{12}$}
\end{subfigure}
\caption{Uniaxial loading problem. $P_{11}$ (a-d) and $P_{12}$ (e-h) stress components.}\label{fig:Prob1P11}
\end{figure}

In a first example we study the displacement and stress results from PINN, DEM, mDEM and PINN for a simple uniaxial loading problem as seen in Figure \ref{fig::UniProbImage}. A $1 \text{m} \times 1 \text{m}$ block is subjected to a uniform line load on the right-hand side while the movement on the left-hand side is restricted. 
The FEM solution is generated from a $100 \times 100$ elements uniform mesh.
The displacement boundary conditions of PINN, DEM and mDEM are a-priori fulfilled by setting
\begin{equation}
    \hat{\bm{u}}(\bm{X}, \bm{\theta}) = \bm{X} \circ \bm{z}(\bm{X}, \bm{\theta}).
\end{equation}
All the methods are trained using a grid of $200 \times 200$ training points over the computational domain (see Figure \ref{fig::UniTrainPoints}).
The boundary traction boundary conditions of PINN and mDEM are trained with an additional 5000 training points per edge.
Even though the loading is uniaxial, as the left side of the domain is clamped we are expecting stress concentrations around the top and bottom of that side.
After training the networks, the displacement magnitudes of DEM, PINN and mDEM follow the expected pattern providing a relatively good match to the FEM result, as displayed in Figure \ref{fig:Prob1DispMagnitude}.
However the $P_{11}$ and $P_{12}$ stress components approximated by the methods show significant differences, see Figure \ref{fig:Prob1P11}. It is clear that mDEM is able to resolve the stresses that arise near the clamping whereas DEM and PINN fail in doing so.

\clearpage
\subsection{Localized traction boundary}
\begin{figure}
\centering
\includegraphics[scale=0.32]{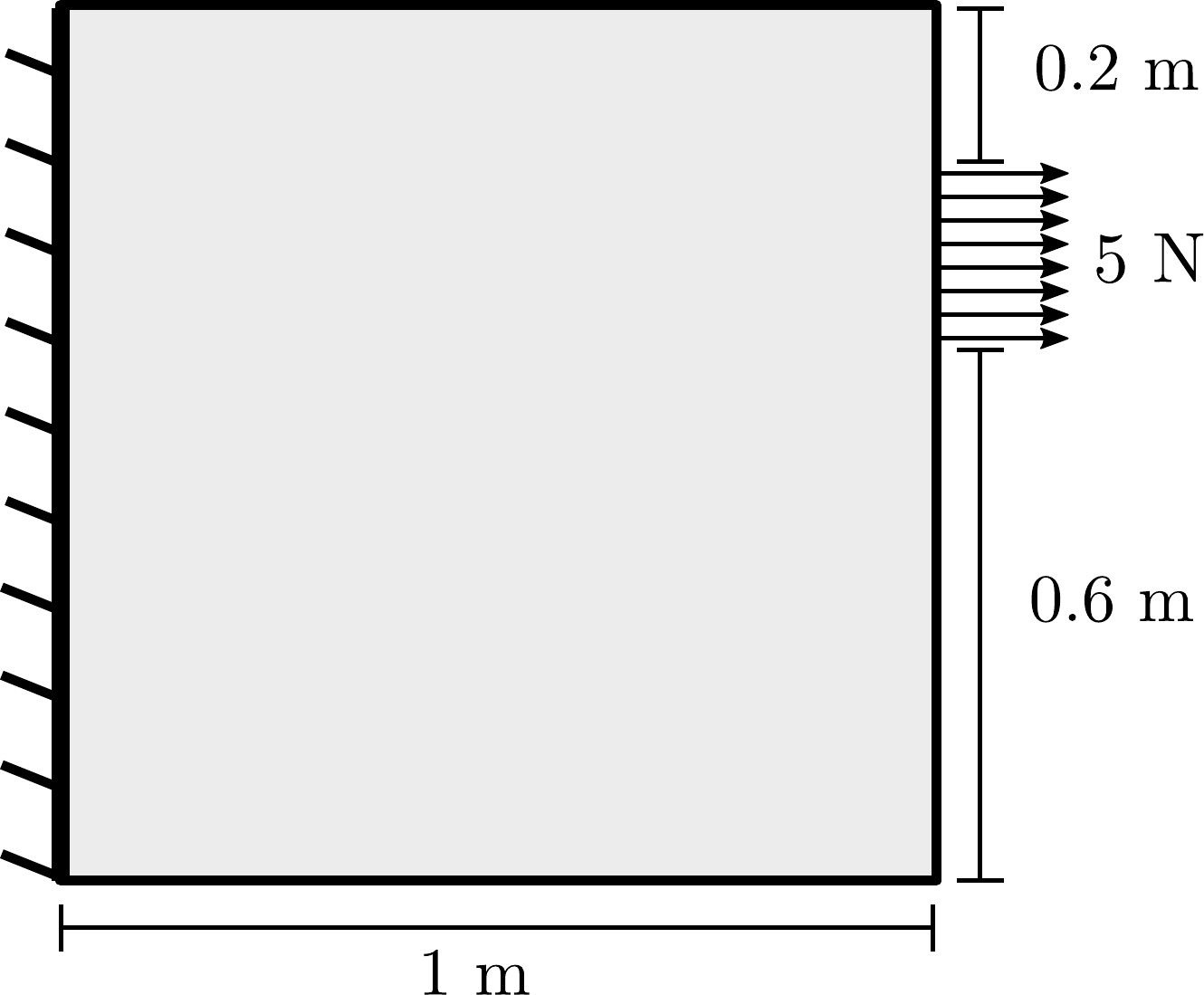}
\caption{Localized traction problem}\label{fig::LocalTractoin}
\end{figure}

\begin{figure}[b!]
\begin{subfigure}[b]{0.5\linewidth}
\centering
\includegraphics[scale=0.18]{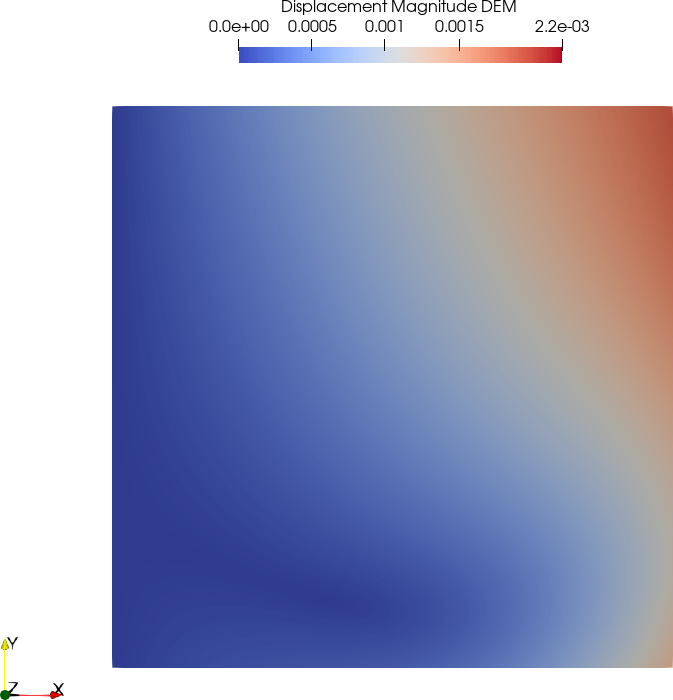} 
\caption{DEM}
\end{subfigure}%
\begin{subfigure}[b]{.5\linewidth}
\centering
\includegraphics[scale=0.18]{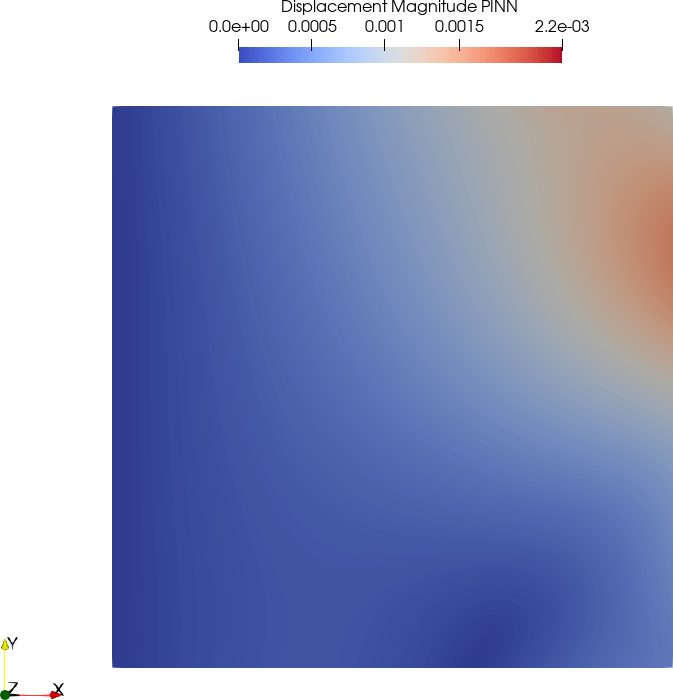} 
\caption{PINN}
\end{subfigure}

\begin{subfigure}[b]{0.5\linewidth}
\centering
\includegraphics[scale=0.18]{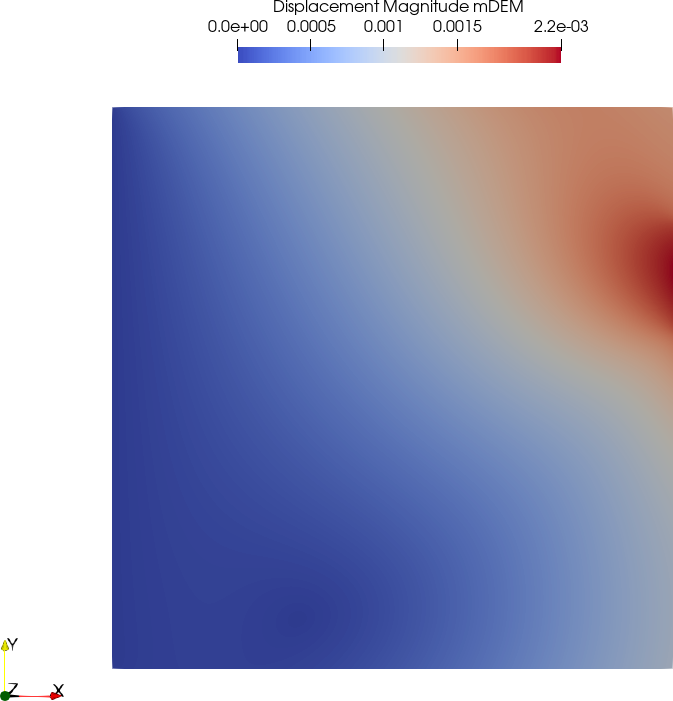} 
\caption{mDEM}
\end{subfigure}%
\begin{subfigure}[b]{0.5\linewidth}
\centering
\includegraphics[scale=0.18]{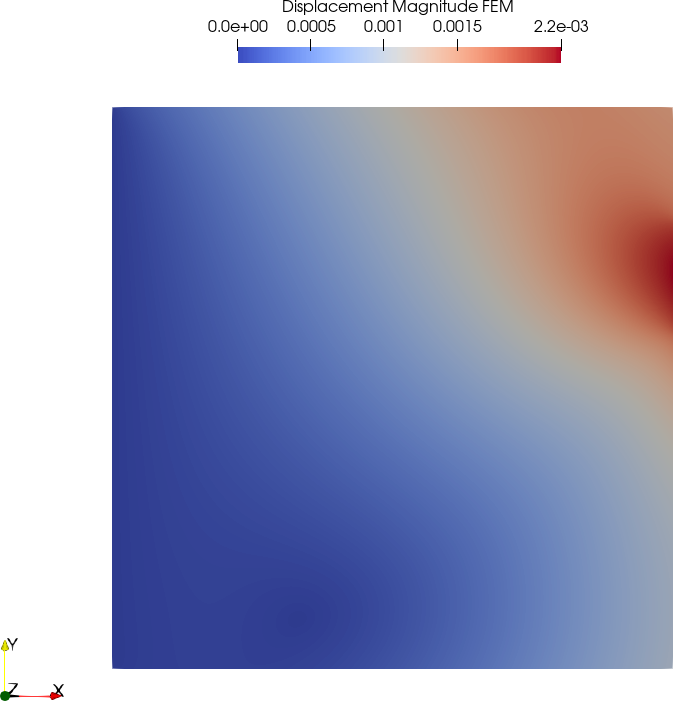} 
\caption{FEM}
\end{subfigure}
\caption{Localized Uniaxial loading problem displacement magnitude.}\label{fig:Prob2UMag}
\end{figure}

\begin{figure}
\begin{subfigure}[b]{0.5\linewidth}
\centering
\includegraphics[scale=0.2]{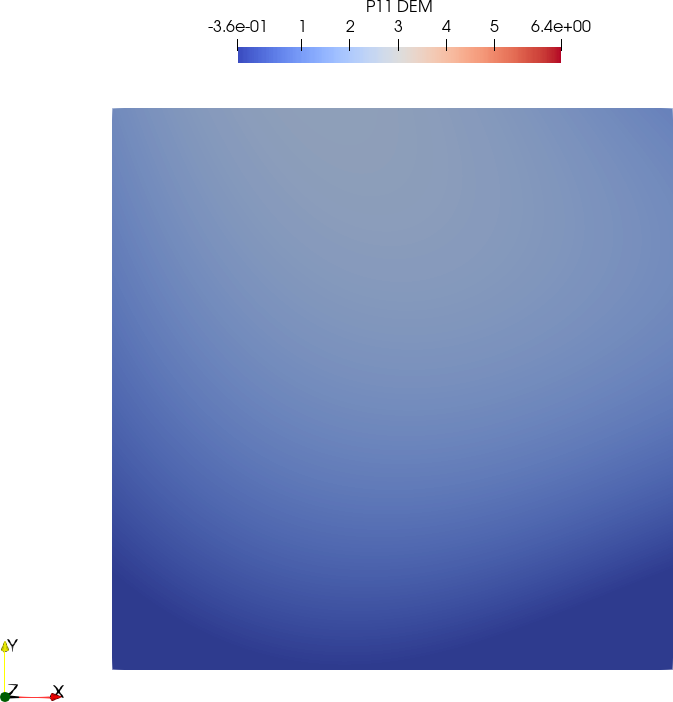} 
\caption{DEM $P_{11}$}
\end{subfigure}%
\begin{subfigure}[b]{.5\linewidth}
\centering
\includegraphics[scale=0.2]{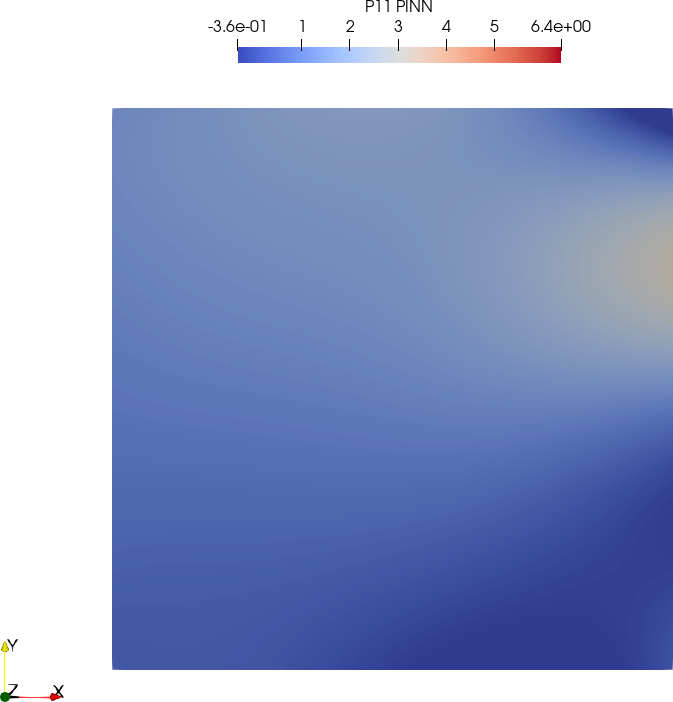} 
\caption{PINN $P_{11}$}
\end{subfigure}

\begin{subfigure}[b]{0.5\linewidth}
\centering
\includegraphics[scale=0.2]{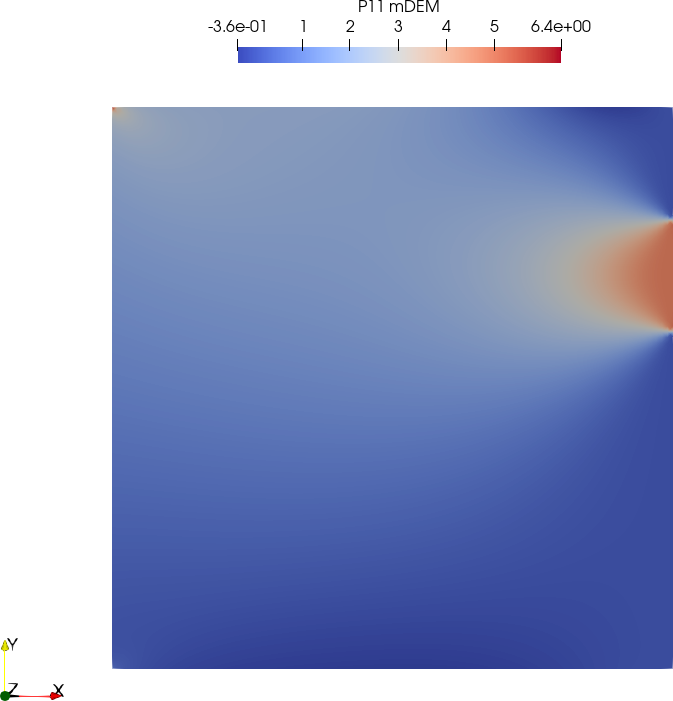} 
\caption{mDEM $P_{11}$}
\end{subfigure}%
\begin{subfigure}[b]{0.5\linewidth}
\centering
\includegraphics[scale=0.2]{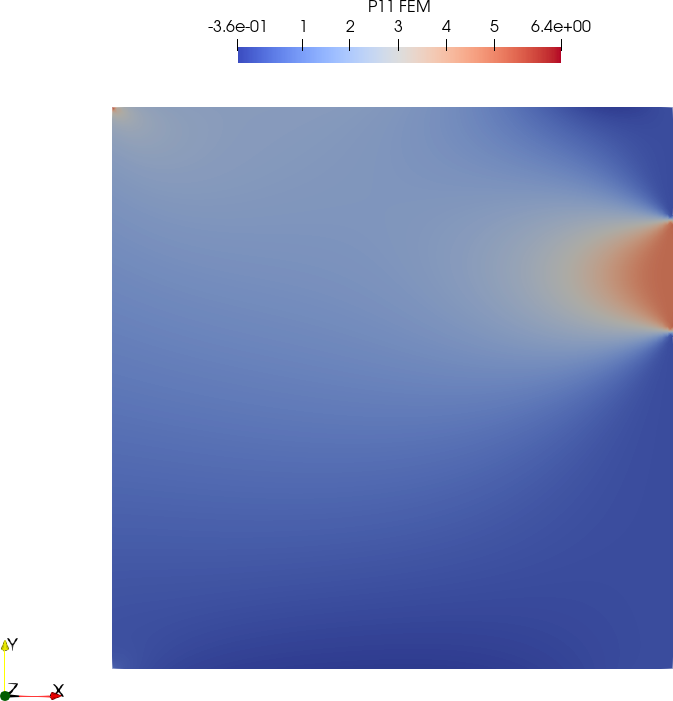} 
\caption{FEM $P_{11}$}
\end{subfigure}

\begin{subfigure}[b]{0.5\linewidth}
\centering
\includegraphics[scale=0.2]{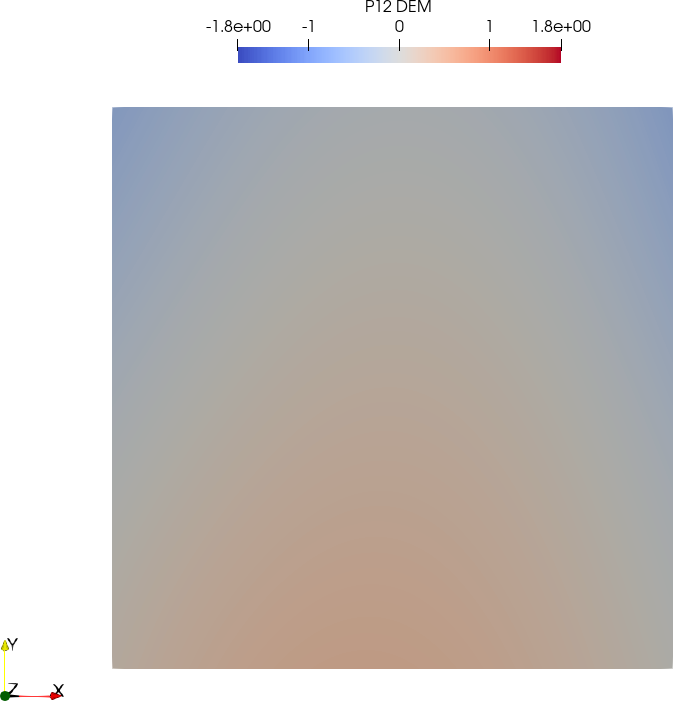} 
\caption{DEM $P_{12}$}
\end{subfigure}%
\begin{subfigure}[b]{.5\linewidth}
\centering
\includegraphics[scale=0.2]{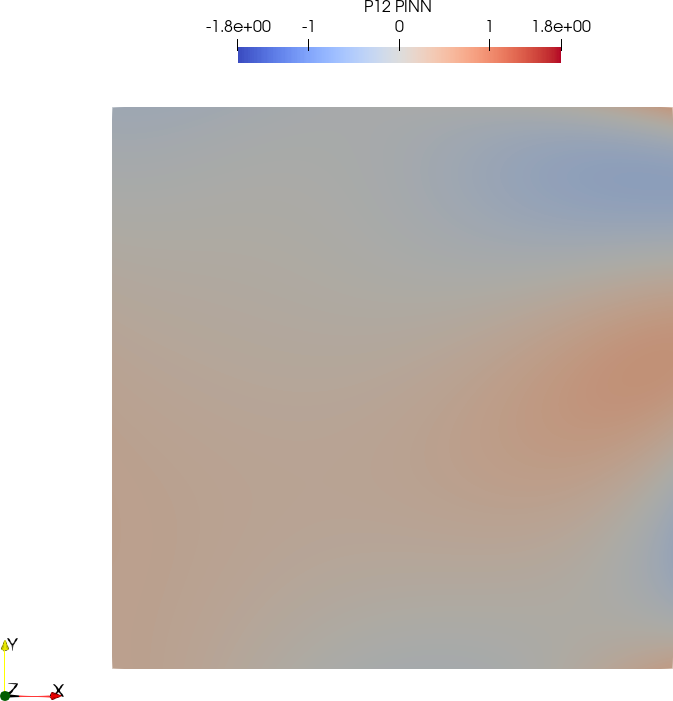} 
\caption{PINN $P_{12}$}
\end{subfigure}

\begin{subfigure}[b]{0.5\linewidth}
\centering
\includegraphics[scale=0.2]{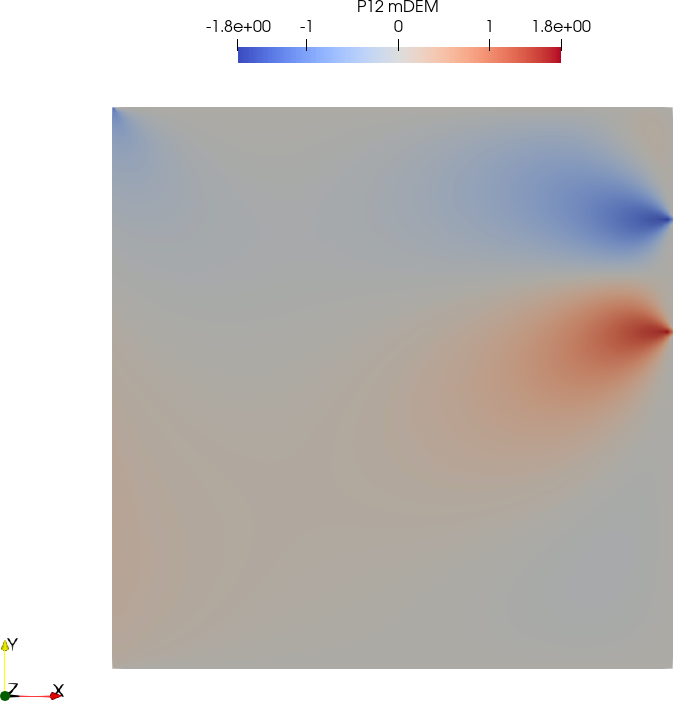} 
\caption{mDEM $P_{12}$}
\end{subfigure}%
\begin{subfigure}[b]{0.5\linewidth}
\centering
\includegraphics[scale=0.2]{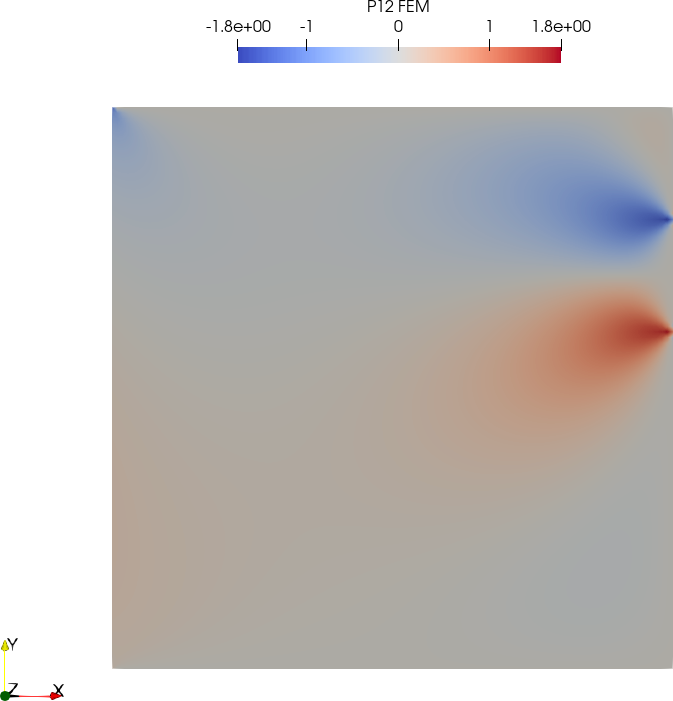} 
\caption{FEM $P_{12}$}
\end{subfigure}
\caption{Localized uniaxial loading problem stress components $P_{11}$ (a-d) and $P_{12}$ (e-f).}\label{fig:Prob2P11}
\end{figure}

The first example considers the $1m \times 1m$ block as depicted in Figure \ref{fig::LocalTractoin}. A  load is applied locally on the right-hand side of the block.
The same number of domain and boundary training points are used as in the previous example (see Section \ref{sec::UniaxialTension} and Figure \ref{fig::UniTrainPoints}). 
The displacement boundary conditions of PINN, DEM and mDEM are a-priori fulfilled by setting
\begin{equation}
    \hat{\bm{u}}(\bm{X}, \bm{\theta}) = \bm{X} \circ \bm{z}(\bm{X}, \bm{\theta}).
\end{equation}
Figure \ref{fig:Prob2UMag} displays the resulting displacement magnitudes of the four different numerical techniques. From the FEM solution it can be seen that there is a displacement concentration around the locally applied traction boundary condition. Comparing DEM, PINN and mDEM results shows that only mDEM is able to resolve this feature accurately.
Figure \ref{fig::ErrorPINNStressConc} shows the error evolution over the training iterations for PINN. It can be seen that even though the error converges sufficiently, PINN is not able to recover the fine features of the FEM solution, which might be due to the stiffness of problem, i.e. requiring an accurate traction boundary fit on a first order automatic differentiation level while needing to resolve the residual at a second order level.
Similarly, the same problems with DEM and PINN can be observed when looking at the resulting stress components $P_{11}$ and $P_{12}$ (Figure \ref{fig:Prob2P11}).
In contrast, mDEM is able to achieve proficient stress fields compared to FEM.

\clearpage

\subsection{Beam with a circular hole}
\begin{figure}[hbtp]
\centering
\begin{subfigure}[b]{0.5\linewidth}
\centering
\includegraphics[scale=0.45]{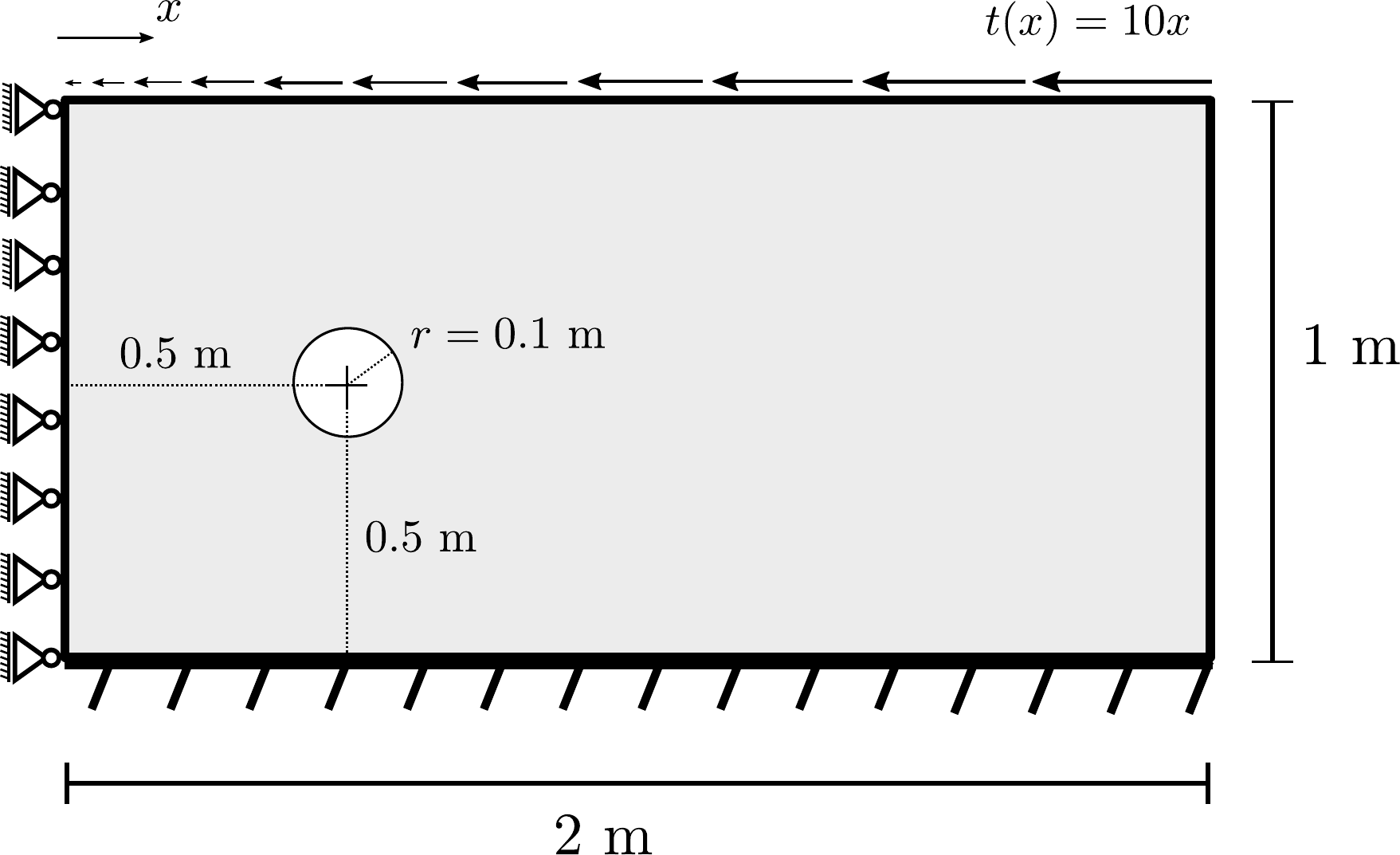}
\caption{Beam with defection}\label{fig::Prob3}
\end{subfigure}%
\begin{subfigure}[b]{0.5\linewidth}
\centering
\includegraphics[scale=0.33]{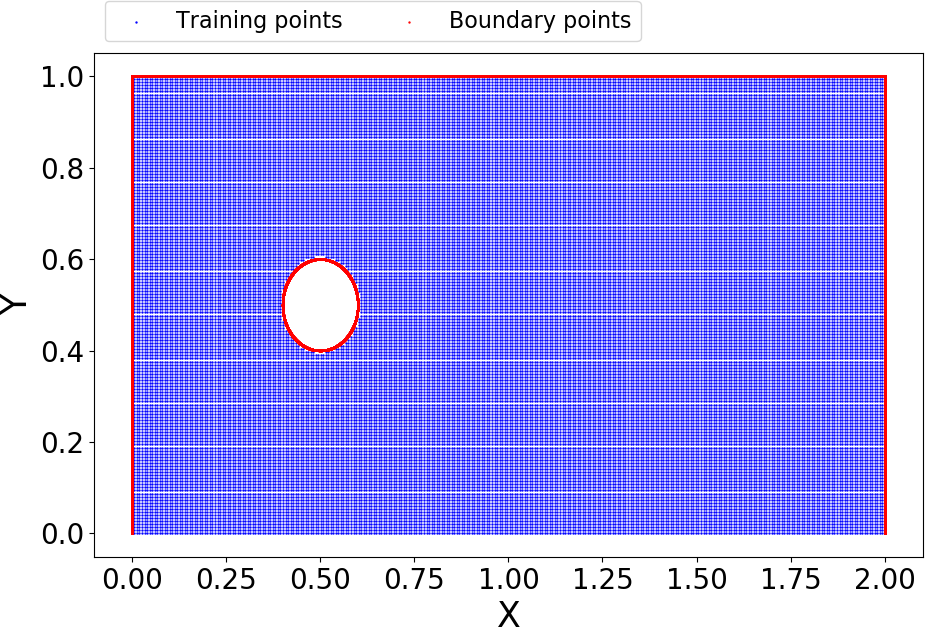}
\caption{Training point distribution}\label{fig::Prob3TrainingPoints}
\end{subfigure}%
\caption{Problem setting and training point positions}
\end{figure}

In the last example we look at a 2D beam with a circular hole, see Figure \ref{fig::Prob3}. On each edge $5000$ training points for the traction boundary were sampled, whereas the domain training points are arranged in a point-grid of $300 \times 150$ samples where points lying in the hole were removed. The final training point positions are displayed in Figure \ref{fig::Prob3TrainingPoints}.
\begin{figure}[h]
\centering
\begin{subfigure}[b]{0.5\linewidth}
\centering
\includegraphics[scale=0.35]{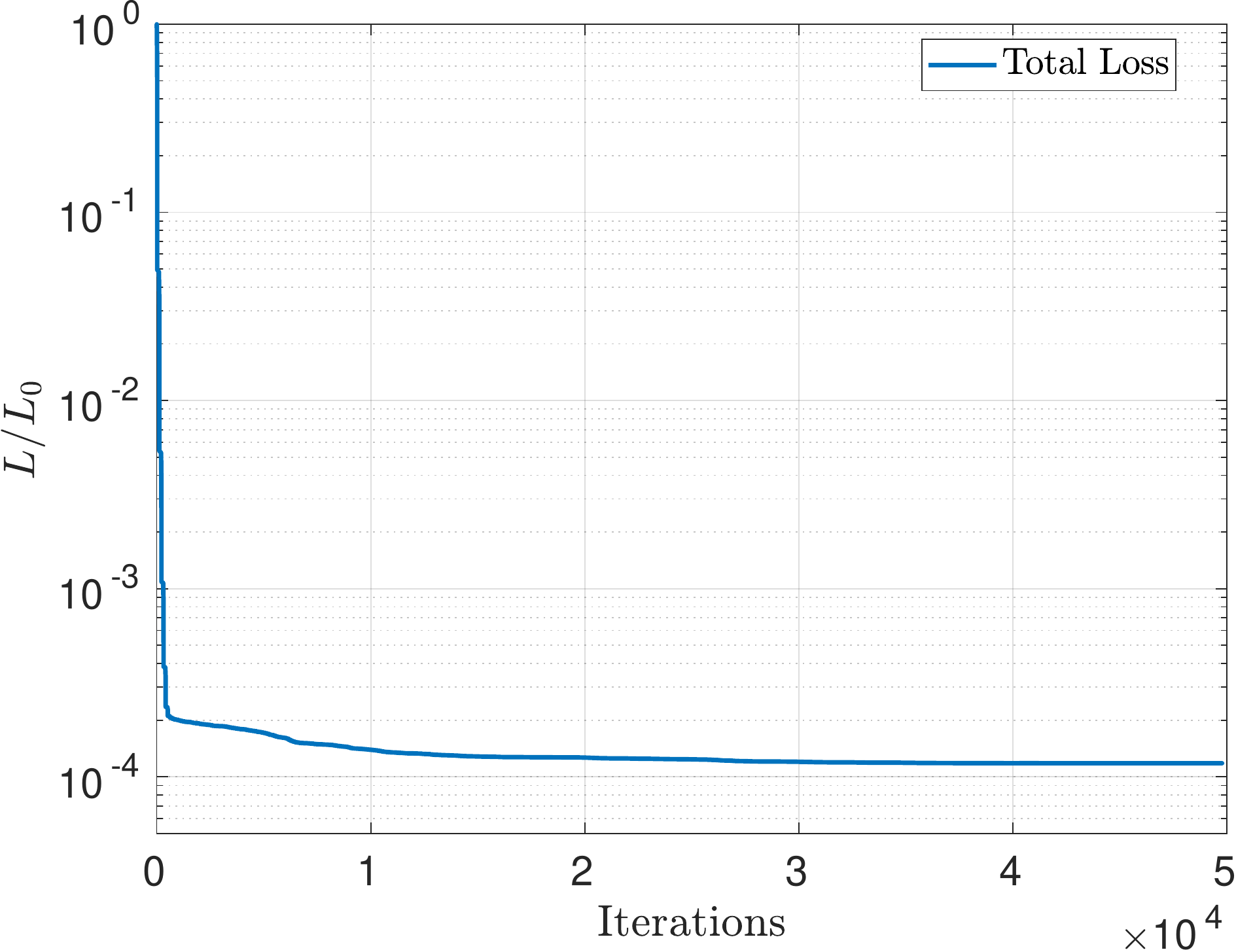}
\caption{Local traction boundary}\label{fig::ErrorPINNStressConc}
\end{subfigure}%
\begin{subfigure}[b]{0.5\linewidth}
\centering
\includegraphics[scale=0.35]{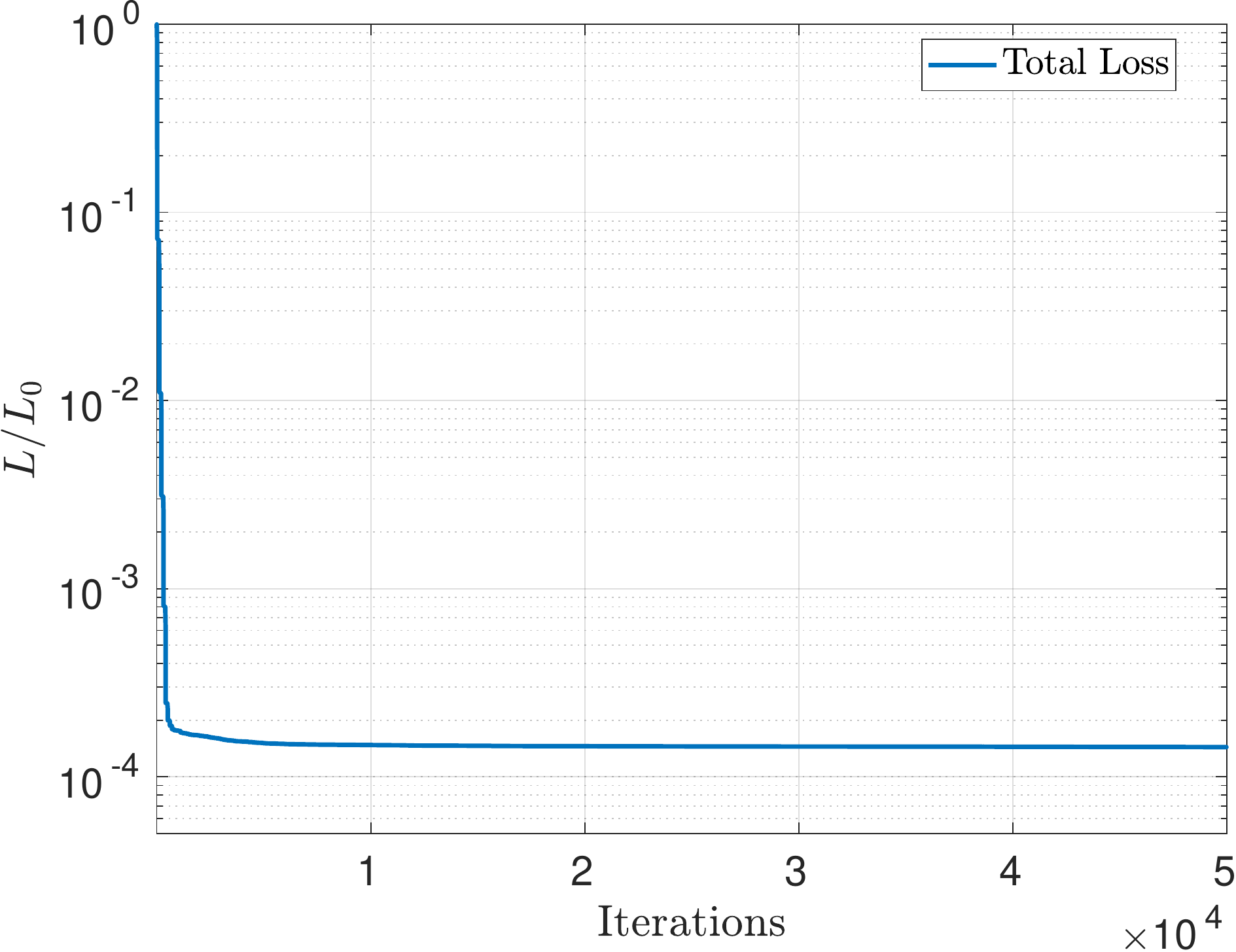}
\caption{Beam with a circular hole}\label{fig::ErrorPINNHole}
\end{subfigure}%
\caption{Normalized loss value convergence over iteration steps of PINN.}\label{fig::}
\end{figure}

\begin{figure}
\begin{subfigure}[b]{0.5\linewidth}
\centering
\includegraphics[scale=0.2]{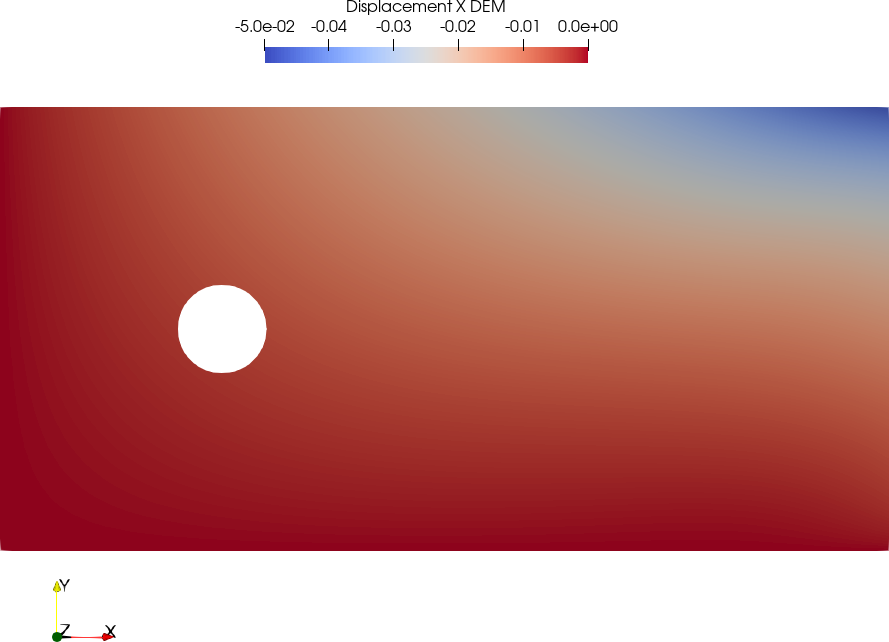} 
\caption{DEM $u_{x}$}
\end{subfigure}%
\begin{subfigure}[b]{.5\linewidth}
\centering
\includegraphics[scale=0.2]{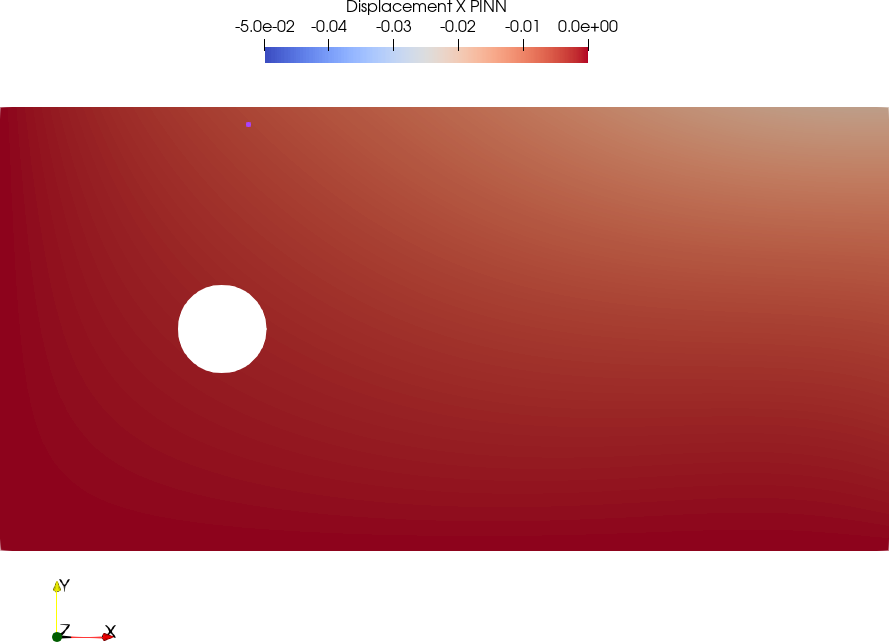} 
\caption{PINN $u_{x}$}
\end{subfigure}

\begin{subfigure}[b]{0.5\linewidth}
\centering
\includegraphics[scale=0.2]{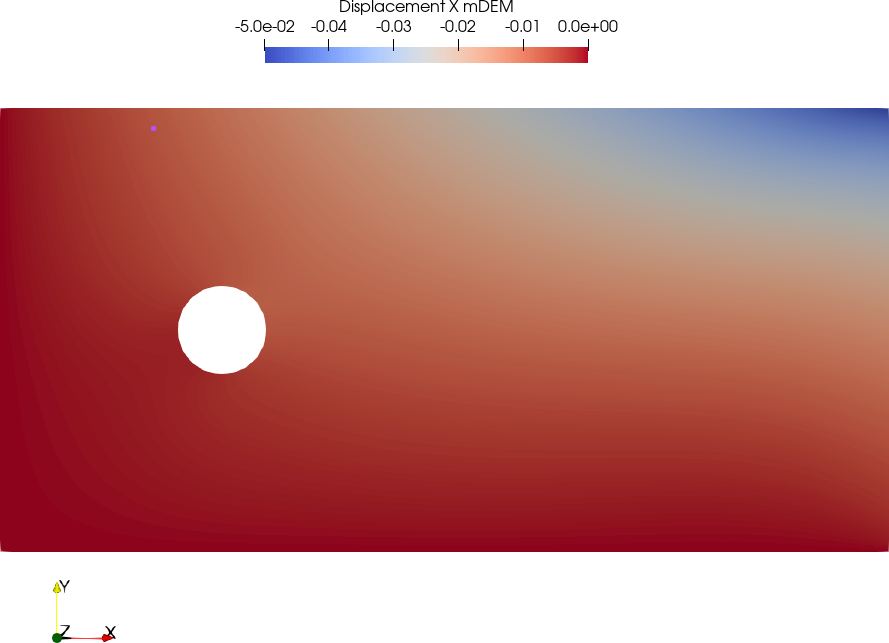} 
\caption{mDEM $u_{x}$}
\end{subfigure}%
\begin{subfigure}[b]{0.5\linewidth}
\centering
\includegraphics[scale=0.2]{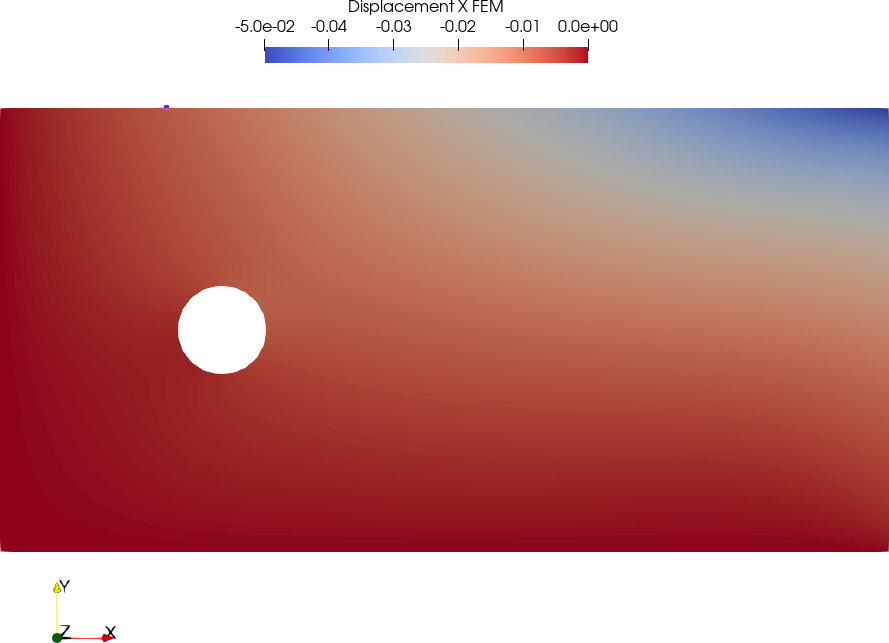} 
\caption{FEM $u_{x}$}
\end{subfigure}

\begin{subfigure}[b]{0.5\linewidth}
\centering
\includegraphics[scale=0.2]{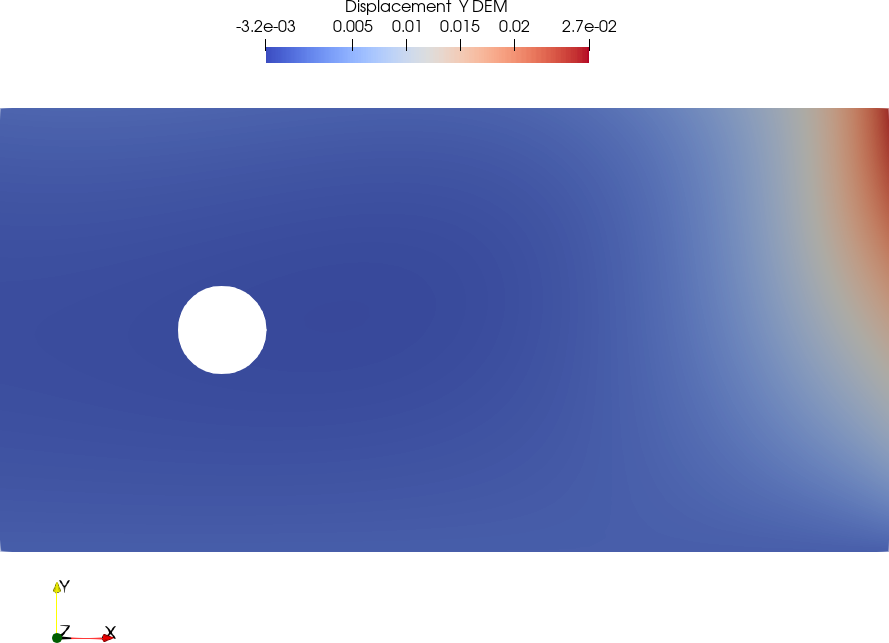} 
\caption{DEM $u_{y}$}
\end{subfigure}%
\begin{subfigure}[b]{.5\linewidth}
\centering
\includegraphics[scale=0.2]{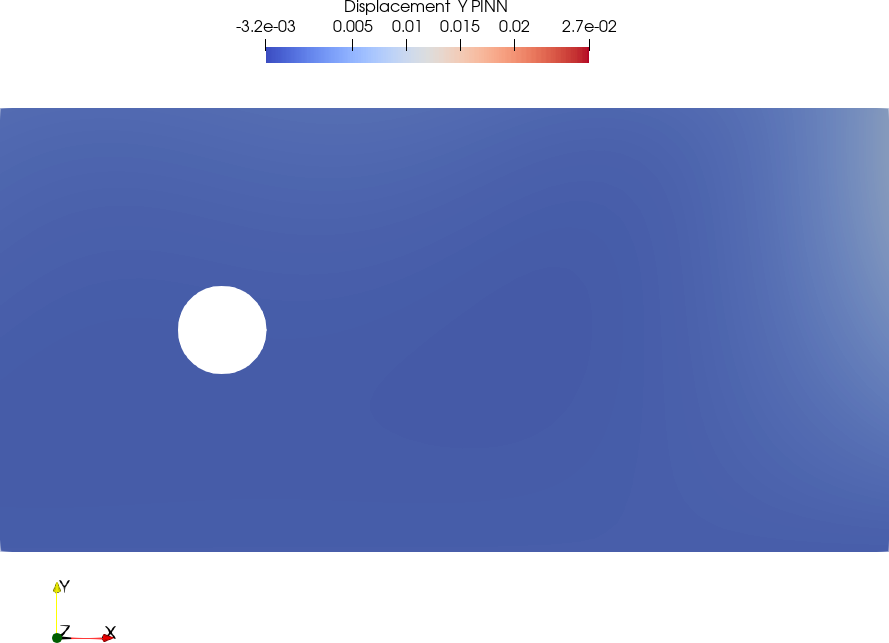} 
\caption{PINN $u_{y}$}
\end{subfigure}

\begin{subfigure}[b]{0.5\linewidth}
\centering
\includegraphics[scale=0.2]{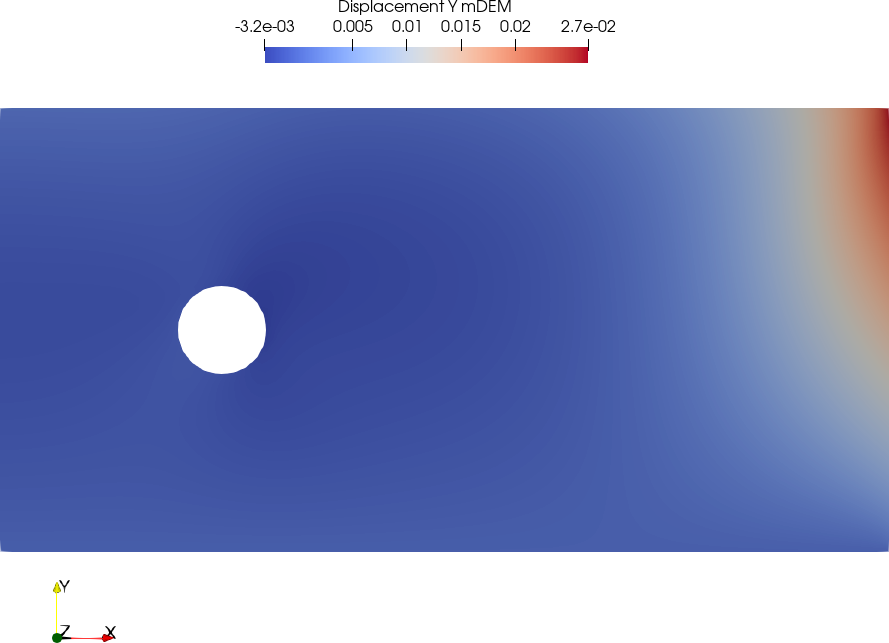} 
\caption{mDEM $u_{y}$}
\end{subfigure}%
\begin{subfigure}[b]{0.5\linewidth}
\centering
\includegraphics[scale=0.2]{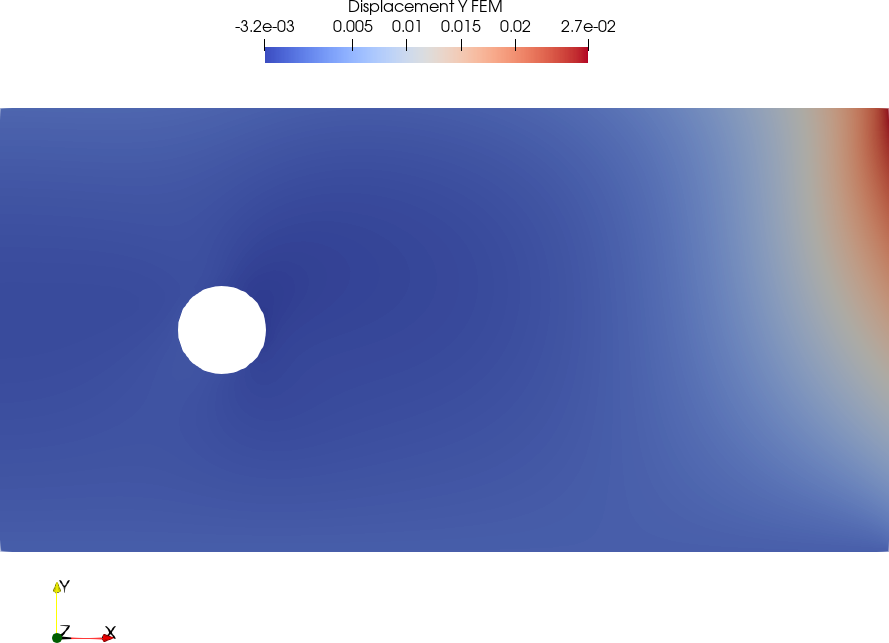} 
\caption{FEM $u_{y}$}
\end{subfigure}
\caption{Defection problem displacement components  $u_{x}$ (a-d) and $u_{y}$ (e-f) .}\label{fig:Prob3ux}
\end{figure}

\begin{figure}
\begin{subfigure}[b]{0.5\linewidth}
\centering
\includegraphics[scale=0.2]{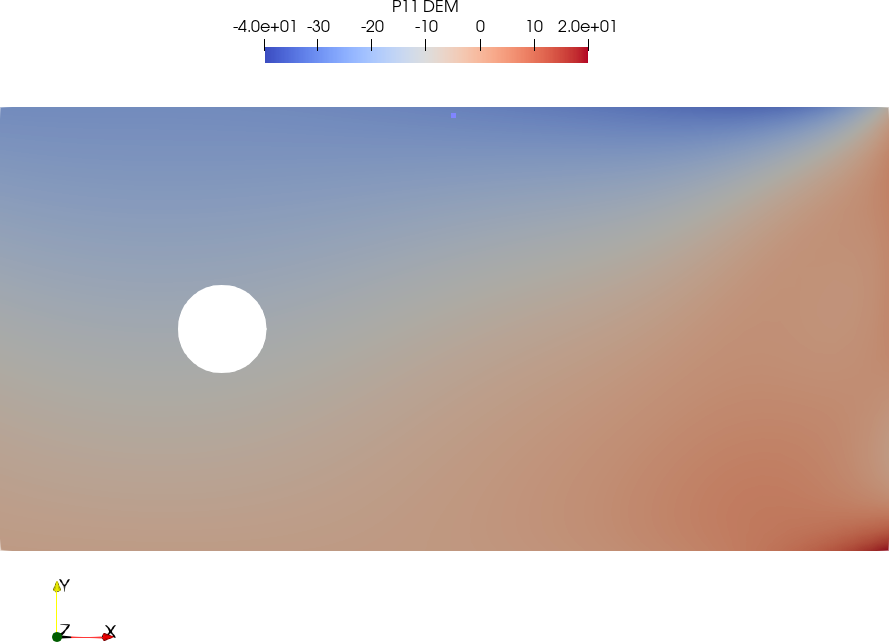} 
\caption{DEM $P_{11}$}
\end{subfigure}%
\begin{subfigure}[b]{.5\linewidth}
\centering
\includegraphics[scale=0.2]{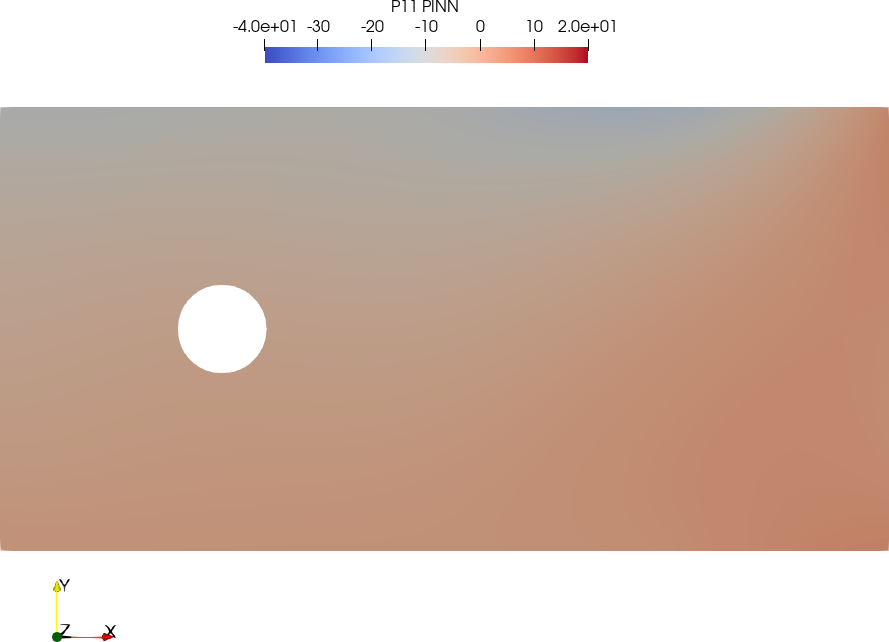} 
\caption{PINN $P_{11}$}
\end{subfigure}

\begin{subfigure}[b]{0.5\linewidth}
\centering
\includegraphics[scale=0.2]{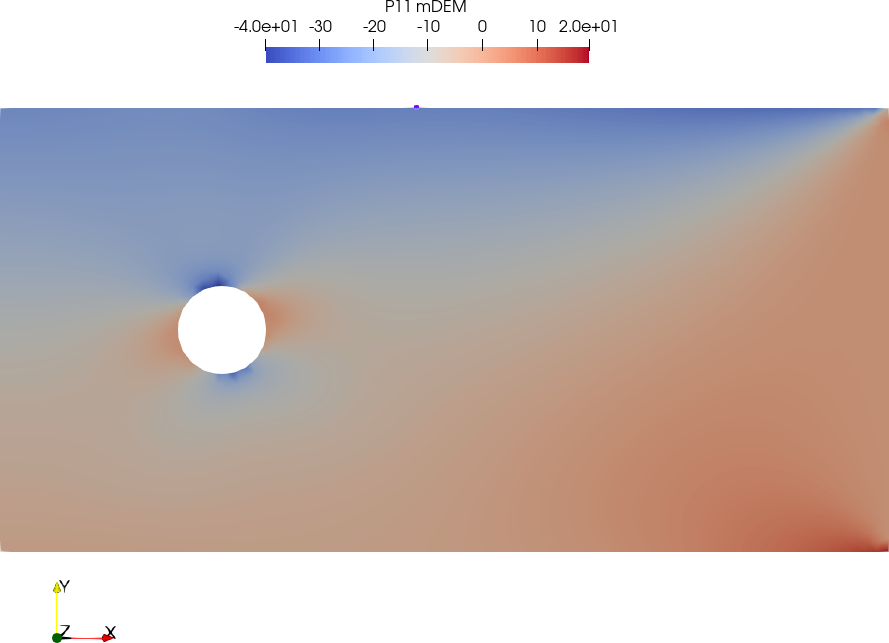} 
\caption{mDEM $P_{11}$}
\end{subfigure}%
\begin{subfigure}[b]{0.5\linewidth}
\centering
\includegraphics[scale=0.2]{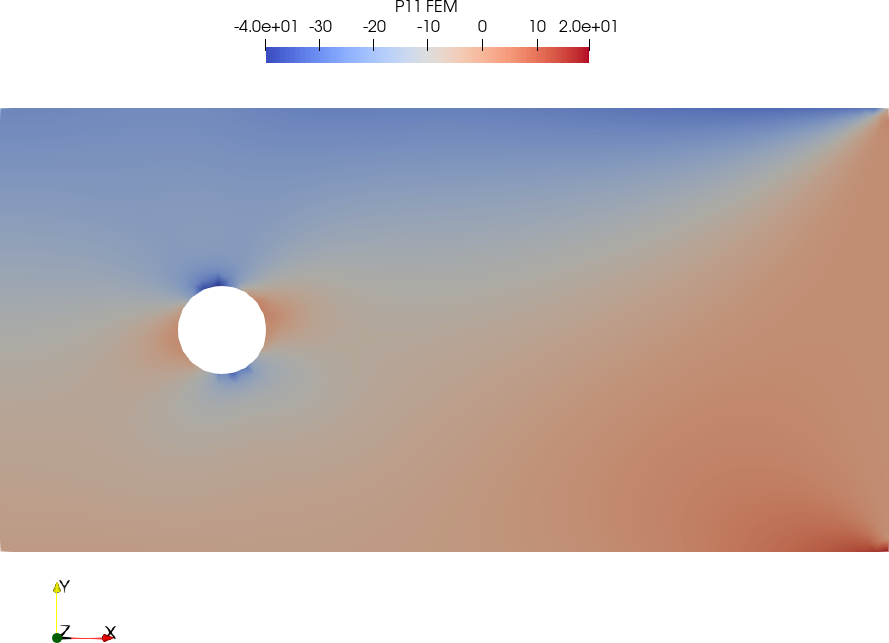} 
\caption{FEM $P_{11}$}
\end{subfigure}

\begin{subfigure}[b]{0.5\linewidth}
\centering
\includegraphics[scale=0.2]{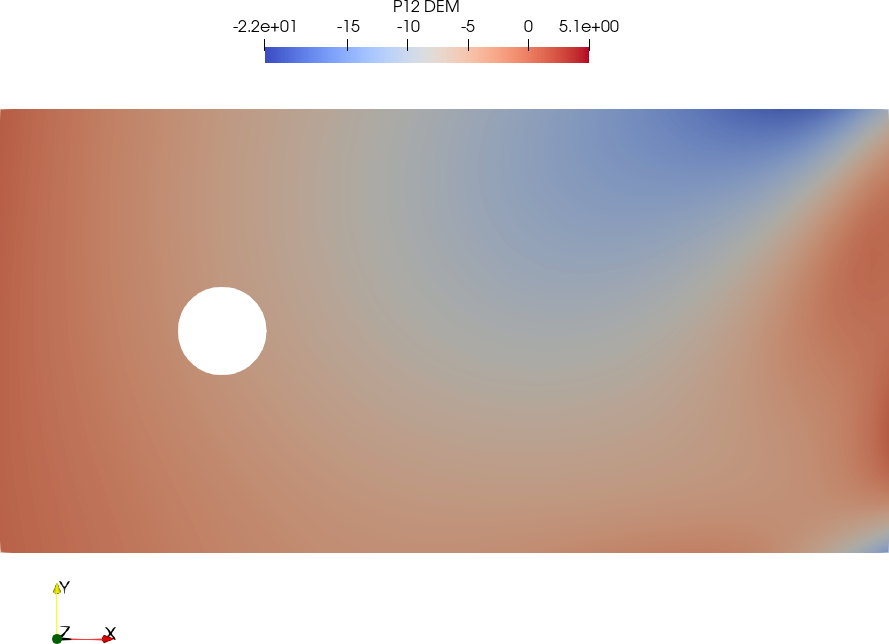} 
\caption{DEM $P_{12}$}
\end{subfigure}%
\begin{subfigure}[b]{.5\linewidth}
\centering
\includegraphics[scale=0.2]{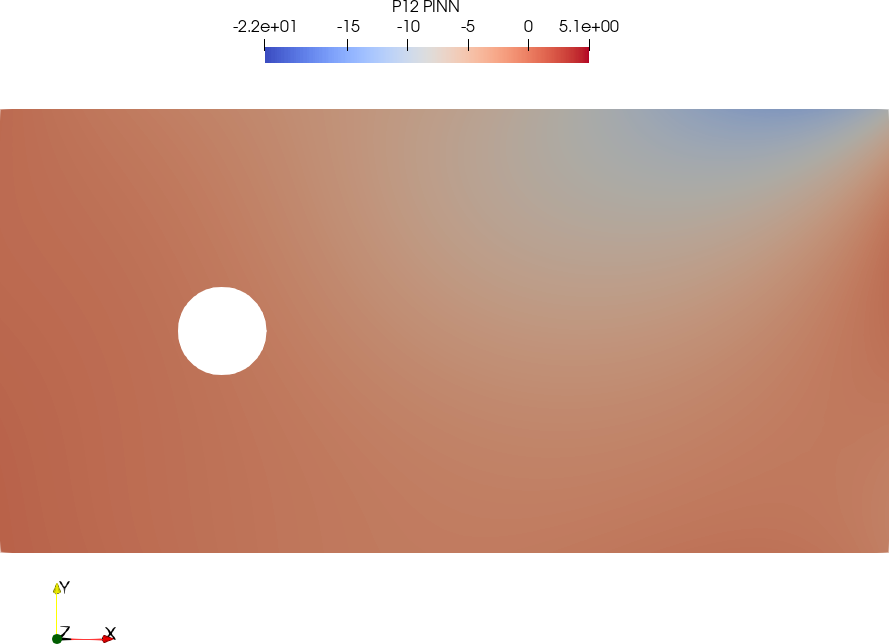} 
\caption{PINN $P_{12}$}
\end{subfigure}

\begin{subfigure}[b]{0.5\linewidth}
\centering
\includegraphics[scale=0.2]{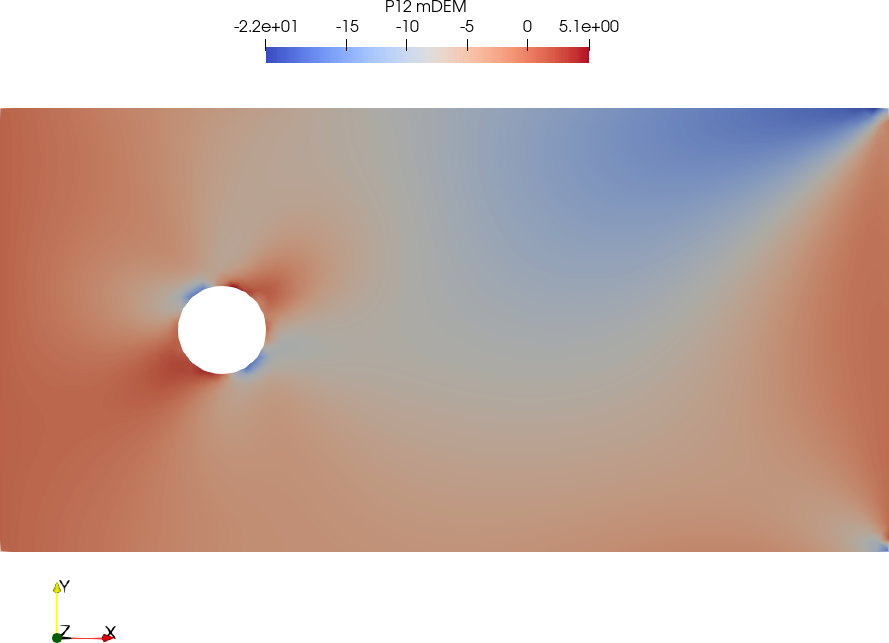} 
\caption{mDEM $P_{12}$}
\end{subfigure}%
\begin{subfigure}[b]{0.5\linewidth}
\centering
\includegraphics[scale=0.2]{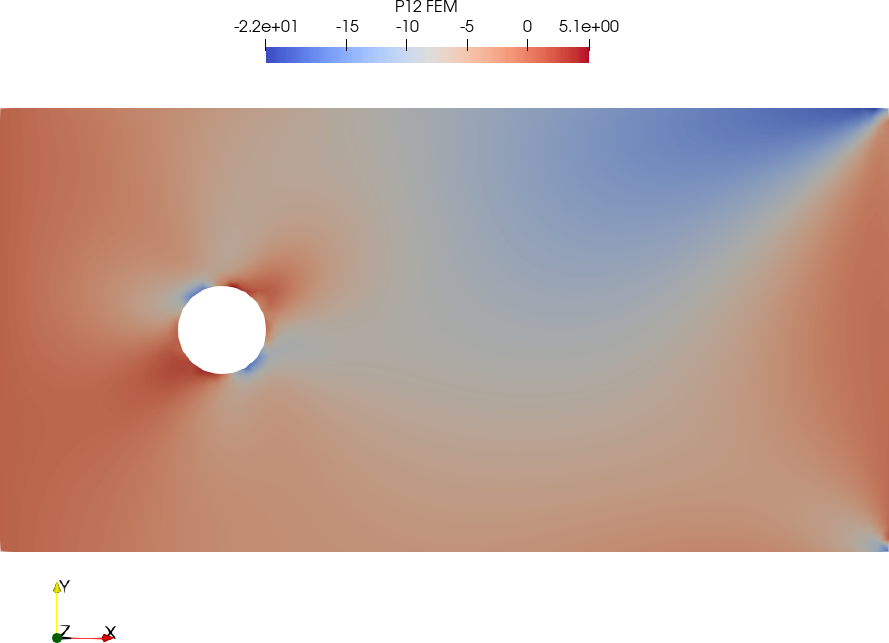} 
\caption{FEM $P_{12}$}
\end{subfigure}
\caption{Defection problem stress components $P_{11}$ (a-d) and $P_{12}$ (e-f).}\label{fig:Prob3P11}
\end{figure}

The displacement boundary conditions of PINN, DEM and mDEM are a-priori fulfilled by defining
\begin{equation}
\begin{aligned}
        \bm{u}_{1}(\bm{X}, \bm{\theta}) &= X Y z_{1}(\bm{X}, \bm{\theta}), \\
        \bm{u}_{2}(\bm{X}, \bm{\theta}) &= Y z_{2}(\bm{X}, \bm{\theta}).
\end{aligned}
\end{equation}
The displacement solutions in the $x-$ and $y-$ directions of all investigated techniques are shown in 
Figure \ref{fig:Prob3ux} respectively. DEM and mDEM are in close agreement to the FEM solutions, whereas PINN does not appear to find the correct solution. This is supported by the evolution of the loss values over the training procedure as reported in 
Figure \ref{fig::ErrorPINNHole} which highlights a quick saturation of the error value pointing towards stagnation in local minima.
When looking at the expected stress concentration features, here displayed for the $P_{11}$ and $P_{12}$ components in Figure
\ref{fig:Prob3P11}, it can be seen that even though DEM is able to follow the overall stress field in an accurate manner, only mDEM resolves the stress concentrations on comparable levels to FEM.
\clearpage
\section{Conclusion and outlook}\label{sec::conclusion}
This paper proposes an extension to the formulation of the Deep Energy Method (DEM) called mixed Deep Energy Method (mDEM).
It enhances the original approach which uses a neural network as a global shape function of the displacement over the computational domain by additionally defining the stress components of the first Piola-Kirchhoff stress tensor as outputs of the networks. This proves to be useful in combating the shortcomings of DEM with regards to resolving concentration features. 
In order to make mDEM more versatile we introduce a numerical integration approach based on Delaunay tesselation that does not require the training point positions to be arranged in grid-like fashion. \\
We test the proposed approach on three computational experiments which exhibit stress concentrations and compare the results to DEM, a Physics-Informed Neural Network and FEM solutions. It can be seen that mDEM is able to yield comparable results to FEM in all investigated problems, whereas DEM and PINN prove to not be reliably able to resolve local stress and displacement features. In future works we plan to apply the presented formulation to three dimensional applications, time-dependent computational experiments as well as problems governed by non-elastic constitutive laws as well as for inverse problems focusing on parameter estimation. 
\clearpage
\bibliography{bib.bib}

\begin{thebibliography}{25}
\expandafter\ifx\csname natexlab\endcsname\relax\def\natexlab#1{#1}\fi
\providecommand{\url}[1]{\texttt{#1}}
\providecommand{\href}[2]{#2}
\providecommand{\path}[1]{#1}
\providecommand{\DOIprefix}{doi:}
\providecommand{\ArXivprefix}{arXiv:}
\providecommand{\URLprefix}{URL: }
\providecommand{\Pubmedprefix}{pmid:}
\providecommand{\doi}[1]{\href{http://dx.doi.org/#1}{\path{#1}}}
\providecommand{\Pubmed}[1]{\href{pmid:#1}{\path{#1}}}
\providecommand{\bibinfo}[2]{#2}
\ifx\xfnm\relax \def\xfnm[#1]{\unskip,\space#1}\fi
\bibitem[{Abueidda et~al.(2020)Abueidda, Lu and Koric}]{abueidda2020deep}
\bibinfo{author}{Abueidda, D.W.}, \bibinfo{author}{Lu, Q.},
  \bibinfo{author}{Koric, S.}, \bibinfo{year}{2020}.
\newblock \bibinfo{title}{Deep learning collocation method for solid mechanics:
  Linear elasticity, hyperelasticity, and plasticity as examples}.
\newblock \bibinfo{journal}{arXiv preprint arXiv:2012.01547} .
\bibitem[{Aln{\ae}s et~al.(2015)Aln{\ae}s, Blechta, Hake, Johansson, Kehlet,
  Logg, Richardson, Ring, Rognes and Wells}]{AlnaesBlechta2015a}
\bibinfo{author}{Aln{\ae}s, M.S.}, \bibinfo{author}{Blechta, J.},
  \bibinfo{author}{Hake, J.}, \bibinfo{author}{Johansson, A.},
  \bibinfo{author}{Kehlet, B.}, \bibinfo{author}{Logg, A.},
  \bibinfo{author}{Richardson, C.}, \bibinfo{author}{Ring, J.},
  \bibinfo{author}{Rognes, M.E.}, \bibinfo{author}{Wells, G.N.},
  \bibinfo{year}{2015}.
\newblock \bibinfo{title}{The fenics project version 1.5}.
\newblock \bibinfo{journal}{Archive of Numerical Software} \bibinfo{volume}{3}.
\newblock \DOIprefix\doi{10.11588/ans.2015.100.20553}.
\bibitem[{Bottou et~al.(2018)Bottou, Curtis and
  Nocedal}]{bottou2018optimization}
\bibinfo{author}{Bottou, L.}, \bibinfo{author}{Curtis, F.E.},
  \bibinfo{author}{Nocedal, J.}, \bibinfo{year}{2018}.
\newblock \bibinfo{title}{Optimization methods for large-scale machine
  learning}.
\newblock \bibinfo{journal}{Siam Review} \bibinfo{volume}{60},
  \bibinfo{pages}{223--311}.
\bibitem[{Eggersmann et~al.(2019)Eggersmann, Kirchdoerfer, Reese, Stainier and
  Ortiz}]{eggersmann2019model}
\bibinfo{author}{Eggersmann, R.}, \bibinfo{author}{Kirchdoerfer, T.},
  \bibinfo{author}{Reese, S.}, \bibinfo{author}{Stainier, L.},
  \bibinfo{author}{Ortiz, M.}, \bibinfo{year}{2019}.
\newblock \bibinfo{title}{Model-free data-driven inelasticity}.
\newblock \bibinfo{journal}{Computer Methods in Applied Mechanics and
  Engineering} \bibinfo{volume}{350}, \bibinfo{pages}{81--99}.
\bibitem[{Fuhg et~al.(2021)Fuhg, Boehm, Bouklas, Fau, Wriggers and
  Marino}]{fuhg2021modeldatadriven}
\bibinfo{author}{Fuhg, J.N.}, \bibinfo{author}{Boehm, C.},
  \bibinfo{author}{Bouklas, N.}, \bibinfo{author}{Fau, A.},
  \bibinfo{author}{Wriggers, P.}, \bibinfo{author}{Marino, M.},
  \bibinfo{year}{2021}.
\newblock \bibinfo{title}{Model-data-driven constitutive responses: application
  to a multiscale computational framework}.
\newblock \href{http://arxiv.org/abs/2104.02650}{{\tt arXiv:2104.02650}}.
\bibitem[{Gonz{\'a}lez et~al.(2019)Gonz{\'a}lez, Chinesta and
  Cueto}]{gonzalez2019thermodynamically}
\bibinfo{author}{Gonz{\'a}lez, D.}, \bibinfo{author}{Chinesta, F.},
  \bibinfo{author}{Cueto, E.}, \bibinfo{year}{2019}.
\newblock \bibinfo{title}{Thermodynamically consistent data-driven
  computational mechanics}.
\newblock \bibinfo{journal}{Continuum Mechanics and Thermodynamics}
  \bibinfo{volume}{31}, \bibinfo{pages}{239--253}.
\bibitem[{Goodfellow et~al.(2016)Goodfellow, Bengio, Courville and
  Bengio}]{goodfellow2016deep}
\bibinfo{author}{Goodfellow, I.}, \bibinfo{author}{Bengio, Y.},
  \bibinfo{author}{Courville, A.}, \bibinfo{author}{Bengio, Y.},
  \bibinfo{year}{2016}.
\newblock \bibinfo{title}{Deep learning}. volume~\bibinfo{volume}{1}.
\newblock \bibinfo{publisher}{MIT press Cambridge}.
\bibitem[{Haghighat et~al.(2020)Haghighat, Raissi, Moure, Gomez and
  Juanes}]{haghighat2020deep}
\bibinfo{author}{Haghighat, E.}, \bibinfo{author}{Raissi, M.},
  \bibinfo{author}{Moure, A.}, \bibinfo{author}{Gomez, H.},
  \bibinfo{author}{Juanes, R.}, \bibinfo{year}{2020}.
\newblock \bibinfo{title}{A deep learning framework for solution and discovery
  in solid mechanics}.
\newblock \bibinfo{journal}{arXiv preprint arXiv:2003.02751} .
\bibitem[{Hernandez et~al.(2021)Hernandez, Bad{\'\i}as, Gonz{\'a}lez, Chinesta
  and Cueto}]{hernandez2021deep}
\bibinfo{author}{Hernandez, Q.}, \bibinfo{author}{Bad{\'\i}as, A.},
  \bibinfo{author}{Gonz{\'a}lez, D.}, \bibinfo{author}{Chinesta, F.},
  \bibinfo{author}{Cueto, E.}, \bibinfo{year}{2021}.
\newblock \bibinfo{title}{Deep learning of thermodynamics-aware reduced-order
  models from data}.
\newblock \bibinfo{journal}{Computer Methods in Applied Mechanics and
  Engineering} \bibinfo{volume}{379}, \bibinfo{pages}{113763}.
\bibitem[{Hornik et~al.(1989)Hornik, Stinchcombe and
  White}]{hornik1989multilayer}
\bibinfo{author}{Hornik, K.}, \bibinfo{author}{Stinchcombe, M.},
  \bibinfo{author}{White, H.}, \bibinfo{year}{1989}.
\newblock \bibinfo{title}{Multilayer feedforward networks are universal
  approximators}.
\newblock \bibinfo{journal}{Neural networks} \bibinfo{volume}{2},
  \bibinfo{pages}{359--366}.
\bibitem[{Huang et~al.(2020)Huang, Fuhg, Wei{\ss}enfels and
  Wriggers}]{huang2020machine}
\bibinfo{author}{Huang, D.}, \bibinfo{author}{Fuhg, J.N.},
  \bibinfo{author}{Wei{\ss}enfels, C.}, \bibinfo{author}{Wriggers, P.},
  \bibinfo{year}{2020}.
\newblock \bibinfo{title}{A machine learning based plasticity model using
  proper orthogonal decomposition}.
\newblock \bibinfo{journal}{Computer Methods in Applied Mechanics and
  Engineering} \bibinfo{volume}{365}, \bibinfo{pages}{113008}.
\bibitem[{Iba{\~n}ez et~al.(2017)Iba{\~n}ez, Borzacchiello, Aguado,
  Abisset-Chavanne, Cueto, Ladeveze and Chinesta}]{ibanez2017data}
\bibinfo{author}{Iba{\~n}ez, R.}, \bibinfo{author}{Borzacchiello, D.},
  \bibinfo{author}{Aguado, J.V.}, \bibinfo{author}{Abisset-Chavanne, E.},
  \bibinfo{author}{Cueto, E.}, \bibinfo{author}{Ladeveze, P.},
  \bibinfo{author}{Chinesta, F.}, \bibinfo{year}{2017}.
\newblock \bibinfo{title}{Data-driven non-linear elasticity: constitutive
  manifold construction and problem discretization}.
\newblock \bibinfo{journal}{Computational Mechanics} \bibinfo{volume}{60},
  \bibinfo{pages}{813--826}.
\bibitem[{Jagtap and Karniadakis(2020)}]{jagtap2020extended}
\bibinfo{author}{Jagtap, A.D.}, \bibinfo{author}{Karniadakis, G.E.},
  \bibinfo{year}{2020}.
\newblock \bibinfo{title}{Extended physics-informed neural networks (xpinns): A
  generalized space-time domain decomposition based deep learning framework for
  nonlinear partial differential equations}.
\newblock \bibinfo{journal}{Communications in Computational Physics}
  \bibinfo{volume}{28}, \bibinfo{pages}{2002--2041}.
\bibitem[{Kadeethum et~al.(2021)Kadeethum, Ballarin and
  Bouklas}]{kadeethum2021non}
\bibinfo{author}{Kadeethum, T.}, \bibinfo{author}{Ballarin, F.},
  \bibinfo{author}{Bouklas, N.}, \bibinfo{year}{2021}.
\newblock \bibinfo{title}{Non-intrusive reduced order modeling of
  poroelasticity of heterogeneous media based on a discontinuous galerkin
  approximation}.
\newblock \bibinfo{journal}{arXiv preprint arXiv:2101.11810} .
\bibitem[{Kadeethum et~al.(2020)Kadeethum, J{\o}rgensen and
  Nick}]{kadeethum2020physics}
\bibinfo{author}{Kadeethum, T.}, \bibinfo{author}{J{\o}rgensen, T.M.},
  \bibinfo{author}{Nick, H.M.}, \bibinfo{year}{2020}.
\newblock \bibinfo{title}{Physics-informed neural networks for solving
  nonlinear diffusivity and biot’s equations}.
\newblock \bibinfo{journal}{PloS one} \bibinfo{volume}{15},
  \bibinfo{pages}{e0232683}.
\bibitem[{Kharazmi et~al.(2021)Kharazmi, Zhang and
  Karniadakis}]{kharazmi2021hp}
\bibinfo{author}{Kharazmi, E.}, \bibinfo{author}{Zhang, Z.},
  \bibinfo{author}{Karniadakis, G.E.}, \bibinfo{year}{2021}.
\newblock \bibinfo{title}{hp-vpinns: Variational physics-informed neural
  networks with domain decomposition}.
\newblock \bibinfo{journal}{Computer Methods in Applied Mechanics and
  Engineering} \bibinfo{volume}{374}, \bibinfo{pages}{113547}.
\bibitem[{Kingma and Ba(2014)}]{kingma2014adam}
\bibinfo{author}{Kingma, D.P.}, \bibinfo{author}{Ba, J.}, \bibinfo{year}{2014}.
\newblock \bibinfo{title}{Adam: A method for stochastic optimization}.
\newblock \bibinfo{journal}{arXiv preprint arXiv:1412.6980} .
\bibitem[{Kirchdoerfer and Ortiz(2016)}]{kirchdoerfer2016data}
\bibinfo{author}{Kirchdoerfer, T.}, \bibinfo{author}{Ortiz, M.},
  \bibinfo{year}{2016}.
\newblock \bibinfo{title}{Data-driven computational mechanics}.
\newblock \bibinfo{journal}{Computer Methods in Applied Mechanics and
  Engineering} \bibinfo{volume}{304}, \bibinfo{pages}{81--101}.
\bibitem[{Lagaris et~al.(1998)Lagaris, Likas and
  Fotiadis}]{lagaris1998artificial}
\bibinfo{author}{Lagaris, I.E.}, \bibinfo{author}{Likas, A.},
  \bibinfo{author}{Fotiadis, D.I.}, \bibinfo{year}{1998}.
\newblock \bibinfo{title}{Artificial neural networks for solving ordinary and
  partial differential equations}.
\newblock \bibinfo{journal}{IEEE transactions on neural networks}
  \bibinfo{volume}{9}, \bibinfo{pages}{987--1000}.
\bibitem[{Nguyen-Thanh et~al.(2020)Nguyen-Thanh, Zhuang and
  Rabczuk}]{nguyen2020deep}
\bibinfo{author}{Nguyen-Thanh, V.M.}, \bibinfo{author}{Zhuang, X.},
  \bibinfo{author}{Rabczuk, T.}, \bibinfo{year}{2020}.
\newblock \bibinfo{title}{A deep energy method for finite deformation
  hyperelasticity}.
\newblock \bibinfo{journal}{European Journal of Mechanics-A/Solids}
  \bibinfo{volume}{80}, \bibinfo{pages}{103874}.
\bibitem[{Paszke et~al.(2019)Paszke, Gross, Massa, Lerer, Bradbury, Chanan,
  Killeen, Lin, Gimelshein, Antiga, Desmaison, Kopf, Yang, DeVito, Raison,
  Tejani, Chilamkurthy, Steiner, Fang, Bai and Chintala}]{NEURIPS2019_9015}
\bibinfo{author}{Paszke, A.}, \bibinfo{author}{Gross, S.},
  \bibinfo{author}{Massa, F.}, \bibinfo{author}{Lerer, A.},
  \bibinfo{author}{Bradbury, J.}, \bibinfo{author}{Chanan, G.},
  \bibinfo{author}{Killeen, T.}, \bibinfo{author}{Lin, Z.},
  \bibinfo{author}{Gimelshein, N.}, \bibinfo{author}{Antiga, L.},
  \bibinfo{author}{Desmaison, A.}, \bibinfo{author}{Kopf, A.},
  \bibinfo{author}{Yang, E.}, \bibinfo{author}{DeVito, Z.},
  \bibinfo{author}{Raison, M.}, \bibinfo{author}{Tejani, A.},
  \bibinfo{author}{Chilamkurthy, S.}, \bibinfo{author}{Steiner, B.},
  \bibinfo{author}{Fang, L.}, \bibinfo{author}{Bai, J.},
  \bibinfo{author}{Chintala, S.}, \bibinfo{year}{2019}.
\newblock \bibinfo{title}{Pytorch: An imperative style, high-performance deep
  learning library}, in: \bibinfo{editor}{Wallach, H.},
  \bibinfo{editor}{Larochelle, H.}, \bibinfo{editor}{Beygelzimer, A.},
  \bibinfo{editor}{d\textquotesingle Alch\'{e}-Buc, F.}, \bibinfo{editor}{Fox,
  E.}, \bibinfo{editor}{Garnett, R.} (Eds.), \bibinfo{booktitle}{Advances in
  Neural Information Processing Systems 32}. \bibinfo{publisher}{Curran
  Associates, Inc.}, pp. \bibinfo{pages}{8024--8035}.
\newblock \URLprefix
  \url{http://papers.neurips.cc/paper/9015-pytorch-an-imperative-style-high-performance-deep-learning-library.pdf}.
\bibitem[{Raissi et~al.(2019)Raissi, Perdikaris and
  Karniadakis}]{raissi2019physics}
\bibinfo{author}{Raissi, M.}, \bibinfo{author}{Perdikaris, P.},
  \bibinfo{author}{Karniadakis, G.E.}, \bibinfo{year}{2019}.
\newblock \bibinfo{title}{Physics-informed neural networks: A deep learning
  framework for solving forward and inverse problems involving nonlinear
  partial differential equations}.
\newblock \bibinfo{journal}{Journal of Computational Physics}
  \bibinfo{volume}{378}, \bibinfo{pages}{686--707}.
\bibitem[{Rao et~al.(2020)Rao, Sun and Liu}]{rao2020physics}
\bibinfo{author}{Rao, C.}, \bibinfo{author}{Sun, H.}, \bibinfo{author}{Liu,
  Y.}, \bibinfo{year}{2020}.
\newblock \bibinfo{title}{Physics informed deep learning for computational
  elastodynamics without labeled data}.
\newblock \bibinfo{journal}{arXiv preprint arXiv:2006.08472} .
\bibitem[{Samaniego et~al.(2020)Samaniego, Anitescu, Goswami, Nguyen-Thanh,
  Guo, Hamdia, Zhuang and Rabczuk}]{samaniego2020energy}
\bibinfo{author}{Samaniego, E.}, \bibinfo{author}{Anitescu, C.},
  \bibinfo{author}{Goswami, S.}, \bibinfo{author}{Nguyen-Thanh, V.M.},
  \bibinfo{author}{Guo, H.}, \bibinfo{author}{Hamdia, K.},
  \bibinfo{author}{Zhuang, X.}, \bibinfo{author}{Rabczuk, T.},
  \bibinfo{year}{2020}.
\newblock \bibinfo{title}{An energy approach to the solution of partial
  differential equations in computational mechanics via machine learning:
  Concepts, implementation and applications}.
\newblock \bibinfo{journal}{Computer Methods in Applied Mechanics and
  Engineering} \bibinfo{volume}{362}, \bibinfo{pages}{112790}.
\bibitem[{Weinan and Yu(2018)}]{weinan2018deep}
\bibinfo{author}{Weinan, E.}, \bibinfo{author}{Yu, B.}, \bibinfo{year}{2018}.
\newblock \bibinfo{title}{The deep ritz method: a deep learning-based numerical
  algorithm for solving variational problems}.
\newblock \bibinfo{journal}{Communications in Mathematics and Statistics}
  \bibinfo{volume}{6}, \bibinfo{pages}{1--12}.

\end{thebibliography}
\end{document}